\documentclass{conm-p-l}

\usepackage{epsf}
\usepackage{amssymb}
\usepackage{amsmath,amsthm}
\usepackage{latexsym}
\usepackage{psfrag}
\usepackage{graphics}
\usepackage{epsfig}
\usepackage{comment,a4wide,enumerate}

\title[Quantum Algorithms]{Quantum Algorithms In Group Theory}

\thanks{The project is supported by EPSRC MathFIT grant GR/87406.}

\author[M.~Batty]{Michael Batty}
\author[S.~L.~Braunstein]{Samuel L. Braunstein}\thanks{The second author currently holds a Royal
Society -- Wolfson Research Merit Award.}
\author[A.~J.~Duncan]{Andrew J. Duncan}
\author[S.~Rees]{Sarah Rees}

\newtheorem{theo}{Theorem}[section]
\newtheorem{prop}[theo]{Proposition}
\newtheorem{lemma}[theo]{Lemma}
\newtheorem{defn}[theo]{Definition}
\newtheorem{exam}[theo]{Example}
\newtheorem{examples}[theo]{Examples}
\newtheorem{exercise}[theo]{Exercise}
\newtheorem{exercises}[theo]{Exercises}

\numberwithin{equation}{section}
\newenvironment{ex}{\begin{exam}\rm}{\end{exam}}

\newenvironment{pf}{\begin{proof}}{\end{proof}}

\newcommand{\nc}{\newcommand}
\def\a{\alpha}
\def\b{\beta}
\def\c{\gamma}
\def\d{\delta}
\def\e{\epsilon}
\def\q{\theta}
\def\w{\omega}
\def\cal{\mathcal}

\nc{\RR}{{\mathbb R}} 
\nc{\ZZ}{{\mathbb Z}} 
\nc{\NN}{{\mathbb N}} 
\nc{\QQ}{{\mathbb Q}} 
\nc{\HH}{{\mathbb H}} 
\nc{\PP}{{\mathbb P}} 
\nc{\CC}{{\mathbb C}} 
\nc{\BB}{{\mathbb B}} 

\nc{\Ima}{\mbox{im}} 
\nc{\im}{\mbox{{\rm Im}}} 
\nc{\Supp}{\mbox{Supp}} 
\nc{\kr}{\mbox{ker}} 
\nc{\coker}{\mbox{coker}} 

\nc{\Sym}{\mbox{Sym}} 
\nc{\GL}{\mbox{GL}} 
\nc{\SL}{\mbox{SL}} 
\nc{\PSL}{\mbox{PSL}} 
\nc{\End}{\mbox{End}} 
\nc{\Aut}{\mbox{Aut}} 
\nc{\Out}{\mbox{Out}} 
\nc{\Inn}{\mbox{Inn}} 
\nc{\Hom}{\mbox{Hom}} 
\nc{\Homeo}{\mbox{Homeo}} 
\nc{\Stab}{\mbox{Stab}} 
\nc{\Orb}{\mbox{Orb}} 
\nc{\id}{\mbox{id}} 

\nc{\Conj}{\mbox{Conj}} 
\nc{\FP}{\mbox{FP}} 

\nc{\val}{\mbox{val}} 
\nc{\str}{\mbox{star}} 
\nc{\Star}{\mbox{Star}} 
\nc{\geom}{\mbox{geom}} 

\nc{\tr}{\mbox{tr}} 
\nc{\Tr}{\mbox{Tr}} 

\nc{\defi}{\mbox{def}} 
\nc{\ep}{\varepsilon} 
\nc{\half}{\frac{1}{2}}
\nc{\bl}{\vspace{2mm}\\} 
\nc{\lte}{\leqslant} 
\nc{\gte}{\geqslant} 
\nc{\ltec}{\preceq} 
\nc{\gtec}{\succeq} 
\nc{\ltc}{\prec} 
\nc{\gtc}{\succ} 
\nc{\pmd}{^{\prime}} 
\nc{\pmmd}{^{\prime\prime}} 
\nc{\ppmd}{^{\prime\prime}} 
\nc{\ra}{\rightarrow} 
\nc{\bks}{\backslash}
\nc{\rank}{\mbox{rank}}
\nc{\Min}{\mbox{Min}} 
\nc{\lcm}{\mbox{lcm}} 
\nc{\modn}{\mbox{ }\mod \mbox{ } n}
\nc{\Hull}{\mbox{Hull}}
\nc{\sgn}{\mbox{sgn}}

\nc{\diff}{\mbox{diff}}
\nc{\slice}{\mbox{slice}}
\nc{\diam}{\mbox{diam}}
\nc{\ce}{\mbox{ce}}
\nc{\st}{\mbox{st}}
\nc{\Ca}{\mbox{Ca}}
\nc{\con}{\mbox{con}}
\nc{\State}{\mbox{State}}
\nc{\DState}{\mbox{DState}}
\nc{\exce}{\mbox{exc}}

\nc{\cone}{\mbox{Con}}
\nc{\Con}{\cone}
\nc{\asdim}{\mbox{asdim}}
\nc{\dis}{\mbox{dis}}
\nc{\Cc}{\mbox{Cc}}
\nc{\PD}{\mbox{PD}}
\nc{\cd}{\mbox{cd}}
\nc{\chr}{\mbox{char}}
\nc{\Diffeos}{\mbox{Diffeos}}
\nc{\Fix}{\mbox{Fix}}
\nc{\Fx}{\mbox{Fix}}
\nc{\Eq}{\mbox{Eq}}
\nc{\Ends}{\mbox{ends}}
\nc{\Vol}{\mbox{Vol}}
\nc{\Mon}{\mbox{Mon}}
\nc{\WCH}{\mbox{WCH}}

\nc{\Iso}{\mbox{Iso}} 
\nc{\Int}{\mbox{Int}}
\nc{\inter}{\mbox{Int}}

\nc{\OS}{\mbox{OS}}
\nc{\IS}{\mbox{IS}}
\nc{\two}{{\bf 2}}
\nc{\tens}{\otimes}
\nc{\btens}{\bigotimes}
\nc{\zvect}{\mathbf{0}}
\nc{\QFT}{{\mathcal Q}}
\nc{\cD}{{\mathcal D}}
\nc{\cG}{{\mathcal G}}
\def \ket#1{\left|#1\right>}
\def \bra#1{\left<#1\right|}
\def \bket#1#2{\left<#1|#2\right>}
\def \kbra#1#2{\left|#1\rangle\langle#2\right|}
\def \set#1{\{#1\}}
\def \quzero {\ket{0}}
\def \quone {\ket{1}}
\def \sgp#1{\left<#1\right>}
\nc{\be}{\begin{enumerate}}
\nc{\ee}{\end{enumerate}}
\nc{\bd}{\begin{description}}
\nc{\ed}{\end{description}}
\def \stref#1{{\bf (\ref{#1})}}
\begin{document}


\maketitle

\begin{abstract}
We present a survey of quantum algorithms, primarily for an intended audience of
pure mathematicians. We place an emphasis on algorithms involving group theory.
\end{abstract}

\tableofcontents

\section{Introduction}
It has been known for some time that simulating quantum mechanics (on a computer based on classical mechanics)
takes time which is exponential in the size of the system. This is because the total quantum state behaves as a
tensor product of the individual states \cite{joszalinden}. In 1982 Feynman \cite{feynman} asked whether we could build a computer based
on the principles of quantum mechanics to facilitate this task of simulation. Deutsch, in 1985, \cite{deutsch} extended the question
and asked whether there are {\em any} problems which can be solved more efficiently on such a quantum computer. He answered the
question in the affirmative, within the abstract setting of black boxes and query complexity. This was done by demonstrating
a property of a black box function which requires two evaluations for its determination on a classical computer, but only one evaluation
on a quantum computer. This work was generalised by Deutsch and Jozsa \cite{deutschjosza} in 1992 with an algorithm
to distinguish between
constant and balanced functions in
a single evaluation. This is an exponential speed-up over the  deterministic  classical case.
However,
use of classical  non--deterministic algorithms removes this exponential gap.

Shor's celebrated algorithm \cite{shor1}, \cite{shor2}
for factoring integers
efficiently on a quantum computer gave the subject a huge boost in 1994, because of its applicability to the
RSA cryptosystem. This factoring is possible due to the
ability of the quantum computer to find the period of a function quickly, given that we are promised in advance that the function is
periodic. It does this using the {\em quantum Fourier transform}, which can be constructed efficiently using a technique similar to
that of the fast Fourier transform. Another well-known quantum algorithm is Grover's algorithm \cite{grover},
which performs an unstructured
search in a list of size $N$ in time $O(\sqrt{N})$. This uses a wholly different approach to Shor's algorithm and is based on the
geometric idea of rotating towards a solution within the quantum state. It is worth noting that a quantum computer can be simulated
on a classical computer (albeit very slowly). So something which is classically algorithmically undecidable (for example the word
problem for finitely presented groups) remains undecidable on a quantum computer. Thus
we only gain improvements in
efficiency: the impact of the model of quantum computing described here is on complexity theory.
We note in passing that  that our  model of quantum computing, though the most generally
accepted is not the only one that has been proposed.
%
For example {\em adiabatic} quantum computing  \cite{farhietal00}, \cite{farhietal01} is based on a continuous version of the usual model,
in which the evolution of the quantum system is given by its Hamiltonian which is dependent
on a parameter which varies smoothly from $0$ to $1$. Interesting methods and questions arise from the
study of this model: see \cite{vandametal}.

What is interesting to us is that group theory is playing an increasingly important part in providing algorithms which are
amenable to quantum computing. The modern setting for the class of quantum algorithms which use the Fourier transform
(including Shor's algorithm and the Deutsch-Jozsa algorithm) is the {\em hidden subgroup problem}. We describe this and mention
the cases where it has and has not been solved. The general case of the hidden subgroup problem
(for an arbitrary finite group) is still open and is known
to include the graph isomorphism problem as a special case.

The main strand of group theory in quantum computing consists of efficient
quantum algorithms in finite groups, of which the hidden subgroup problem is but one.
Results relating to the group non-membership problem are proved by Watrous in
\cite{watrous1}, where he shows that this problem lies in a quantum complexity class analogous to Babai and Moran's
{\em Merlin-Arthur games} class MA (as defined in \cite{babaiamgames}).
Our article culminates with another recent algorithm due to Watrous \cite{watrous2},
which efficiently finds the order of a black box solvable group. This builds on Shor's algorithm but also contains essential new
ingredients which seem to crucially depend on group-theoretic structure.

Sections $2$,$3$,$4$ and $6$ of these notes are based on a course of eight lectures given by the first named author to staff and
postgraduate students
at the University of Newcastle upon Tyne in the summer of 2002. He thanks all those who took part for their interest and enthusiasm.
He also thanks John Watrous for useful conversations at the University of Calgary in March 2003.

Sources we found especially useful while preparing this document were
\cite{braunstein} and
\cite{rieffelpolak} for a first overview of quantum computing, \cite{hirvensalo} and \cite{nielsenchuang} for further depth
and general background,
with \cite{ekertjosza} and \cite{pittenger} providing the more detailed aspects of Shor's algorithm; the former for its exposition of the
number-theoretic aspects in particular, and the latter for its account of implementing the quantum Fourier transform efficiently. It should be noted at this point that we have taken Shor's
approach to this algorithm rather than that of Kitaev \cite{kitaev95}, mainly because the former
came to our attention first.
For the Deutsch-Jozsa algorithm, \cite{ouellette} and \cite{cleveetal} provided good recent accounts. We made use of Jozsa's survey
\cite{joszahs} for the hidden subgroup problem, as well as many sources quoted in
Section \ref{subs:hsp}. Watrous' work was taken from the original sources, although we have changed some of the
notation to make it clearer to ourselves.

We thank the referee for careful reading of the manuscript many helpful remarks.
%

\newpage

\section{The basics of quantum computing}

\subsection{An overview}
%

Before we start to describe the mathematical nuts and bolts of quantum
computing it is perhaps worth describing informally what happens during
a quantum computation, and how it differs from a classical
computation. Later, in Section \ref{subs:postulates}, we shall give a fuller and more formal account
of our standing model of quantum computation.

A quantum system  is both constrained by and enriched by the quantum mechanics
which apply to the physical device used to store and manipulate data. These devices may consist for
example of ion traps, optical photons or nuclear magnetic resonance systems, among others (see \cite{nielsenchuang}
for a brief account of possible technologies).
At this stage the reader may start to feel anxiety through lack of knowledge of
quantum physics,
but in fact no physics is required to understand the computational model
described in this article: such
knowledge is needed only to understand where the seemingly strange rules come from. In any case
we shall cover what's needed as and when required.

The memory (register) of a classical computer consists of  a set of {\em classical bits}, each of which
can be in one of two states, $0$ or $1$. We can therefore view the memory of an $n$--bit computer
as the set $\ZZ^n_2$, the direct sum of $n$ copies of $\ZZ_2=\ZZ/2\ZZ$.
The states of such a computer are binary sequences of
length $n$, which we regard as elements of $\ZZ^n_2$.
Computations then consist of sequences of
functions $f:\ZZ_2^n\ra\ZZ^n_2$, which allow the state of the
system to be transformed.
For example the classical NOT gate is the function $\neg:\ZZ_2\ra \ZZ_2$ given by $\neg (i)=i+1 \mod 2$.
The final
state determines the result of the computation.

The memory of a quantum computer consists of  a finite dimensional complex vector space $V$, with an
inner product (in fact a Hilbert
space).  A state of this quantum computer is  a unit vector in $V$.
Given a set $X$ we denote by $\CC X$ the complex vector space with basis the elements of $X$.
For example $\CC\ZZ_2$ is a $2$--dimensional vector space.
Corresponding to the classical $n$--bit computer above we have the
$n$
 {\em quantum bit} or $n$--{\em qubit} quantum computer which has
memory consisting of the  $2^n$--dimensional vector space
$V=\CC\ZZ_2^{\tens  n}$ (by which we mean the $n$--fold tensor product of $\CC\ZZ_2$).
Note that the dimension of $\CC\ZZ_2^{\tens  n}$ is the same as the dimension
of $\CC(\ZZ_2^n)$: elements of the basis of both vector spaces are in one to one
correspondence with $n$--tuples
of elements of $\ZZ_2$. However the inner product in these spaces is not the same
and we shall see, in due course, that it is the tensor product which best
encapsulates the physical characteristics of quantum mechanics.

A quantum computation consists of three phases, an input phase,
an evolutionary phase and a measurement or output phase.
We assume that we can prepare simple input states, for example states consisting of basis vectors.
Computations then consist of sequences of
unitary linear transformations $\phi:V\ra V$ which alter the state of the quantum computer from
$v$ to $\phi(v)$.
For example, if $\ZZ_2$ has elements $0$ and $1$ we may identify these with
basis vectors $\mathbf e_0$, $\mathbf e_1$ of the vector space $\CC\ZZ_2$.  Then,
corresponding to the
classical NOT gate above, we have the quantum NOT gate: the unitary transformation
$\overline{\neg}:\CC\ZZ_2\ra\CC\ZZ_2$ given by $\overline{\neg}(\a{\mathbf e_0}
+\b{\mathbf e_1})=
\a{\mathbf e_1}+\b{\mathbf e_0}$, for all $\a,\b\in\CC$. That is, $\overline{\neg}$ permutes the
basis of $\CC\ZZ_2$.
Observe that if the computer is in state
$({\mathbf e_0}-{\mathbf e_1})/\sqrt{2}$ then one application of the
quantum NOT gate results in the state
$(-{\mathbf e_0}+{\mathbf e_1})/\sqrt{2}$.
%
This behaviour, and its
higher dimensional analogues, give rise to what is known as {\em quantum parallelism}.

Quantum computing also derives additional power from  the existence of
unitary transformations which, unlike the previous example, do not arise from permutations of the basis of the
quantum system. For example there is a unitary map, of $\CC\ZZ_2$ to itself, sending
basis vectors ${\mathbf e_0}$ and ${\mathbf e_1}$ to
$({\mathbf e_0}+{\mathbf e_1})/\sqrt{2}$ and
$({\mathbf e_0}-{\mathbf e_1})/\sqrt{2}$, respectively. There is no corresponding classical transformation
as the images of the basis vectors cannot be stored on a one--bit classical register.

After a computation on a classical computer the output can be observed and preserved for future use if necessary.
A major difficulty in quantum computation is that the process of observation, or measurement,
of the state of the quantum computer
may,
according to the laws of quantum mechanics, alter the state of the system at the instant of measurement.
That is, measurement entails a transformation of the quantum state and the result
observed is the transformed state.
The process is, however, probabilistic  rather than
non--deterministic.
For instance
if the quantum state is $\a_1v_1+\cdots \a_{m}v_{m}$,
where $\a_i\in\CC$, $\{v_i: i=1,\ldots, m\}$ is an orthonormal basis for
$V$ and $|\a_1|^2+\cdots+|\a_{m}|^2=1$, then, under a standard measurement scheme,
we observe $v_i$ with probability
$|\a_i|^2$.
Typically in a quantum algorithm states are manipulated to change
probabilities so that some characteristic of the output can be detected. In fact most
quantum algorithms are Monte Carlo algorithms which have some probability of success
bounded away from zero. This allows us to make
the probability of failure of the algorithm arbitrarily small, by repetition.

\subsection{Dirac's bra--ket notation and conventions for linear algebra}\label{subs:lin_alg}

We refer the reader to \cite{halmos} as  standard reference for linear algebra and  \cite{kadisonringrose} for further details of linear operators.
We consider only finite dimensional complex vector spaces equipped with inner products.
A {\em quantum state} means a unit vector in such a space. A
{\em superposition}
means a linear combination of given vectors which is of unit length.
An {\em amplitude} is a coefficient of a vector expressed
in terms of some fixed basis.
If we have a superposition of vectors in which all the (non--zero) amplitudes have equal size then
we say we have a {\em uniform} superposition.
These terms are
all standard within the field of quantum computation.

The use of {\em ket} $\ket{\cdot\,\,}$ and {\em bra} $\bra{\cdot\,\,}$ to denote vectors and their
duals is also standard in quantum mechanics. Although we do not use it heavily in this article it is
a concise and manageable notation, once it becomes familiar, so we include a description before moving on
to a detailed discussion of quantum computation.

A  vector, with label $\psi$, is written using the ket notation as $\ket{\psi}$.
For instance, if $X$ is a set then $\{\ket{x}\,:\, x\in X\}$ denotes an orthonormal basis for
$\CC X$, with the usual inner product. Also, if $V$ is an $m$--dimensional inner product space then
we write $\{\quzero, \ldots \ket{m-1}\}$ to denote an orthonormal basis of $V$.
Our convention, which
seems to be the standard in quantum mechanics, is that the inner product on $V$ is linear in the second variable
and conjugate--linear in the first variable. Using the bra--ket notation we write the inner product of
$(\ket{x},\ket{y})\in V\times V$ as $\bket{x}{y}$. If $W$ is also an inner  product space and $T$
is a linear operator from $V$ to $W$ then $\bket{x}{T|y}$ denotes the inner product of $(\ket{x},T(\ket{y})$, where
$x\in W$ and $y\in V$ (and the inner product is that of $W$). There is an immediate pay--off to the use
of this notation as follows. If $V$ has orthonormal basis
$\{\ket{v_1},\ldots,\ket{v_m}\}$ and $W$ has orthonormal basis $\{\ket{w_1},\ldots,\ket{w_n}\}$ then the matrix
of $T$ with respect to these bases is easily seen to have ($i$,$j$)th entry $\bket{w_i}{T|v_j}$.

Given $T$ as above we write $T^{\dagger}$ for the {\em adjoint} of $T$: that is the unique linear operator $T^{\dagger}$ from $W$ to $V$ such that
$\bket{T^\dagger(\ket{x})}{y}=\bket{x}{T|y}$ (where the inner product on the left hand side is that of $V$). If
$A$ is the matrix of $T$ then the matrix of $T^\dagger$ is the conjugate transpose of $A$ and is also denoted
$A^\dagger$.
A linear operator $T$ is {\em unitary} if $T^\dagger T=I=TT^\dagger$.

We may regard a vector $\ket{x}\in V$ as a linear operator from $\CC$ to $V$, taking $1$ to $\ket{x}$. Then
the dual of $\ket{x}$ is $\ket{x}^\dagger$, the linear
functional taking $\ket{y}$ to $\bket{x}{y}$. Using bra notation, we write $\bra{x}$ for $\ket{x}^\dagger$ and
then $\bra{x}(\ket{y})=\bket{x}{y}$.
To reconcile this with the more familiar row and column notation
for vectors and their duals, suppose that $\ket{x}$ and $\ket{y}$ are the column vectors $(1,i,2-i)^T$ and
$(2i,1+i,1)^T$, respectively. Then $\ket{x}^\dagger$ is $(1,-i,2+i)$ and so
\[\bra{x}(\ket{y})=(1,-i,2+i)(2i,1+i,1)^T=
                                           3+2i=\langle (1,i,2-i)^T,(2i,1+i,1)^T\rangle=\bket{x}{y}.\]

Continuing in the spirit of this example, suppose $V$ and $W$ have orthonormal bases $\{v_1,v_2,v_3\}$
and $\{w_1,w_2\}$, respectively, with respect to which
$\ket{x}=(a_1,a_2,a_3)^T$ and $\ket{y}=(b_1,b_2)^T$. Then
we can evaluate
the outer, or tensor, product of $\ket{x}$ and $\ket{y}$ which is given by
$(a_1,a_2,a_3)^T(\overline{b}_1,\overline{b}_2)$. This is the
$3\times 2$ matrix with $(i,j)$th entry $a_i\overline{b}_j=\bket{v_i}{x}\bket{y}{w_j}$.
This suggests that if $V$ and $W$ are vector spaces, $x\in W$ and $y\in V$,
then  $\ket{x}\bra{y}$ be defined as the linear operator
from $V$ to $W$ given the rule
\[ \kbra{x}{y}\left(\ket{v}\right)=
\ket{x}\bket{y}{v}=\bket{y}{v}\ket{x}, \,\textrm{ for }\, v\in V,
\]
that is, $\kbra{x}{y}$ is the tensor product of $\ket{x}$ and $\ket{y}$.
In fact $\kbra{x}{y}\in W^\dagger \tens V$, where $W^\dagger$ denotes the dual space of $W$.

We may also regard $\kbra{x}{y}$ as the bilinear map $V\times W^{\dagger}\ra\CC$ which sends $\kbra{v}{w}$ to
\[
\bket{w}{\left(\kbra{x}{y}\right) | v}=\bket{w}{x}\bket{y}{v} .
\]

For example $\CC\ZZ_2$ has basis $\{\quzero,\quone\}$ and  $\kbra{0}{1}$ is the linear operator sending
$\quone$ to $\quzero$ and $\quzero$ to the zero vector $\zvect$. If we identify $\quzero$ and $\quone$ with
$(1,0)^T$ and $(0,1)^T$, respectively, then  this linear operator
has matrix
\[
\left(
\begin{array}{ll}
0 & 1 \\ 0 & 0
\end{array}
\right)
\]
The Dirac notation for the transformation is more concise than the matrix representation. In fact
the size of matrices required to describe linear maps
grows
exponentially in the number of qubits of the quantum computer.
%

As another example, the quantum NOT gate can be written as
\[ |0\rangle\langle 1|+|1\rangle\langle 0| .\]

Tensor products of vector spaces are always over $\CC$.
If $\ket{v}\in V$ and $\ket{w}\in W$ then $\ket{v}\tens \ket{w}$ may be written
as either $\ket{v}\ket{w}$ or as $\ket{vw}$ and we use all these notations interchangeably,
as convenient.
Moreover this notation extends to $n$--fold tensor products in the obvious way.
For example the tensor product $\CC\ZZ_2\tens \CC\ZZ_2$ is a $4$--dimensional vector space
with basis consisting of $\ket{00}$, $\ket{01}$, $\ket{10}$ and $\ket{11}$.
More generally $\CC\ZZ_2^{\tens n}$ is $2^n$ dimensional and
has a basis consisting of $\ket{0\cdots 0},\ldots, \ket{1\cdots 1}$, which we
can write as $\quzero,\ldots , \ket{2^n-1}$, by identifying $(i_0\cdots i_{n-1})\in \ZZ^{n}_2$ with
the integer $2^{n-1}i_0\times \cdots \times 2^0 i_{n-1}$; that is, by regarding elements of $\ZZ^n_2$ as
binary expansions of integers.

Note that if $M$ and $N$ are complex inner product spaces then
$M\tens N$ can be made into an inner product space by defining
the inner product of $a\tens b$ with $c\tens d$ as
$\langle a\tens b|c\tens d\rangle=\langle a,c\rangle\langle b,d\rangle$.
Thus if $x_i, y_i\in V_i$ and $\ket{x}=\ket{x_1\cdots x_n}$ and $\ket{y}=\ket{y_1\cdots y_n}$
are elements of an $n$--fold tensor product $\tens_{i=1}^nV_i$, then
\begin{equation}\label{e:inner_tens}
\bket{x}{y}=\prod_{i=1}^{n} \bket{x_i}{y_i}.
\end{equation}
 In particular if
$V=\CC\ZZ_2^{\tens n}$ and $\ket{x}$ and $\ket{y}$ are elements of $\ZZ_2^n$ then
\begin{equation}
\bket{x}{y}=\prod_{i=1}^{n} \delta_{x_iy_i}\label{e:tinnerp},
\end{equation}
where $\delta_{ij}$ is the Kronecker delta.

Suppose that $\theta:V_1\ra V_2$ and $\phi:W_1\ra W_2$ are linear transformations.
Then $\theta\tens\phi$ is a linear transformation of $V_1\tens W_1\ra V_2\tens W_2$.
Suppose that $A$ and $B$ are the matrices of $\theta$ and $\phi$ with respect to
some fixed bases of $V_i$ and $W_i$, $i=1,2$. These bases induce bases of
$ V_1\tens W_1$ and $V_2\tens W_2$ in a natural way:
if $v$ and $w$ are basis vectors of $V_1$ and
$W_1$, respectively, then $v\tens w$ is a basis vector of $V\tens W$.
We adopt the convention that these natural induced bases are ordered so that
the matrix $A\tens B$ of $\theta \tens \phi$ is
the {\em right Kronecker product}
\[ \left(
  \begin{array}{lll}
   a_{11}B & \cdots & a_{1m}B \\
   \vdots & & \vdots \\
   a_{n1}B & \cdots & a_{nm}B
  \end{array}
  \right)
\]
of $A$ and $B$.

\subsection{The postulates of quantum mechanics}\label{subs:postulates}
We shall now describe more thoroughly how the physical laws of quantum mechanics give
rise to a model of quantum computation and establish a framework within which the
theory of quantum computation may be developed. Our account is taken largely from
\cite{nielsenchuang}, where further details may be found.

A quantum computer consists of a quantum mechanical system, necessarily isolated from
the surrounding environment, so that its behaviour may be externally controlled and is
not disturbed by events unrelated to the control procedures. The following postulates
provide a model for such a system. Discussion of whether this is the best
or correct model of quantum mechanics is outside the scope of this article.
\renewcommand{\labelenumi}{\theenumi}
\be[{\bf Postulate} \bf 1:]
\item \label{p:1} {\bf Quantum mechanical systems}~\\[.5em]
\hspace*{-.5in}\parbox{.9\textwidth}{Associated to an isolated quantum mechanical system is a
complex inner product space $V$. The state of the system at any time is described by
a unit vector in $V$.}
\ee

As we are concerned with computation using limited resources we consider only finite dimensional
systems. The basic system we shall consider is the $2$-dimensional space $\CC\ZZ_2$ with basis
$\{\quzero, \quone\}$, known as a single {\em qubit} system. A state of a single qubit system
is a vector $\a\quzero +\b\quone$, where $|\a|^2+|\b|^2=1$.  If neither $\a$ nor $\b$ is zero the state
is called a superposition. For example,
$\left(\quzero+\quone\right)/\sqrt 2$ is a superposition, which is also uniform.

We may also consider the $n$--dimensional space   $\CC\ZZ_n$ with basis
$\{\ket{0},\ldots, \ket{n-1}\}$ as a basic quantum system. However it seems likely
that physical implementations of quantum computers will normally be restricted to
systems built up from qubits.
We defer consideration of more complex systems until after
Postulate \ref{p:4}.

Next we consider how transformation of the system from one state to another may be realised: that is
how to program the computer.

\renewcommand{\labelenumi}{\theenumi}
\be[{\bf Postulate }\bf 1:]
\setcounter{enumi}{1}
\item {\bf Evolution}\label{p:2} ~\\[.5em]
\hspace*{-.5in}\parbox{.9\textwidth}{
Evolution of an isolated quantum mechanical system is described by unitary transformations. The states $\ket{v_1}$ and
$\ket{v_2}$ of
the system at times $t_1$ and $t_2$, respectively, are related by a unitary transformation
$\phi$, which depends only on $t_1$ and $t_2$, such that
$\phi(\ket{v_1})=\ket{v_2}$.}
\ee

In the case of quantum computing evolution takes place at discrete intervals of time, finitely often, so
evolution of the system is governed by a finite sequence of unitary transformations.
In the classical setting, elementary computable functions are commonly referred to as {\em gates}.
Analogously, certain basic unitary transformations of a complex vector space are referred to
as {\em quantum gates}. In this article, since we do not wish to become involved in discussion
of the technical details of implementation of unitary transformations on a quantum computer,
we shall refer to any unitary transformation as a gate.

\be[{\bf Postulate }\bf 1:]
\setcounter{enumi}{2}
\item {\bf Measurement}\label{p:3}~\\[.5em]
\hspace*{-.5in}\parbox{.9\textwidth}{
A measurement of  a quantum system consists of a set $\{M_m: m=1,\ldots ,k\}$ of linear
operators on $V$, such that
\begin{equation}\label{e:measurement}
\sum_{m=1}^{k}M_m^\dagger M_m=I.
\end{equation}
The measurement result is one of the indices $m$. If $V$ is in state $\ket{v}$ then the probability that
$m$ observed is
\[p(m)=\bket{v}{M_m^\dagger M_m|v}=\bket{M_m(\ket{v})}{M_m\ket{v}}.\]
If $m$ is observed then the state of $V$ is transformed from $v$ to
\[\frac{M_m\ket{v}}{\sqrt{p(m)}}.\]
}
\ee
Observe that \eqref{e:measurement} implies that $p$ is a probability measure as
\begin{equation*}
\sum_{m=1}^k p(m) =\sum_{m=1}^k\bket{v}{M_m^\dagger M_m|v}=\bket{v}{I|v}=1.
\end{equation*}

We usually restrict attention to the special case of measurement where
$M_m$ is self--adjoint and $M_m^2=M_m$, for all $m$, and $M_mM_n=0$, when $m\neq n$. Such measurements
are called {\em projective} measurements. In the case of projective measurement there are mutually orthogonal
subspaces $P_1,\ldots , P_k$ of $V$ such that $V=\sum_m P_m$ and $M_m=\sum_i\kbra{i}{i}$, for some
orthonormal basis $\{\ket{i}:i=0,\ldots, d_m-1\}$ of $P_m$. That is $M_m$ is projection onto $P_m$.
It turns out, as shown in \cite{nielsenchuang} that, with the use of
Postulates \ref{p:2} and \ref{p:4}, measurements of the type described in
Postulate \ref{p:3} can be achieved using only projective measurements.

In fact most measurements we make will be of the form $\{M_m=\kbra{m}{m}: m=0,\ldots ,n-1\}$, where
$V$ is $n$--dimensional and the basis used to
describe the state vector and evolution of $V$  is $\{\ket{m}:m=0,\ldots ,n-1\}$ . These are called
{\em measurements with respect to the computational basis}. In this article all measurements
are taken with respect to the computational basis, unless they're explicitly defined.

For example suppose we have a single qubit system in state $Q=a\quzero+b\quone$,
where $|a|^2+|b|^2=1$.
If we observe $Q$ with respect to the computational basis we obtain
$0$ with probability $\bket{Q}{0\rangle\bra{0}Q}=|a|^2$ and $1$ with probability
$\bket{Q}{1\rangle\bra{1}Q}=|b|^2$.
The quantum system enters the state
\[\frac{\ket{0}\bket{0}{Q}}{|a|}=\frac{a}{|a|}\quzero,\] if $0$ is measured and
\[\frac{\ket{1}\bket{1}{Q}}{|b|}=\frac{b}{|b|}\quone,\] if $1$ is measured.
It turns out, as we'll see in Section \ref{subs:phase}, that these factors, $a/|a|$ and $b/|b|$, of
modulus $1$ can be ignored and we can assume that the system is either in state
$\quzero$ or $\quone$ after measurement.
Note that if $i$ was measured then
further measurements in the computational basis will result in $i$ with probability $1$.

We shall write
\[p_M(\ket{\psi}\ra a)\]
for the probability that $a$ is observed when a register containing $\ket{\psi}$ is
measured with a measurement $M$. 

We shall often abuse notation by saying that $M_m\ket{v}$ is observed when we mean that $m$ is
observed, as the result of a measurement.  In particular, when measuring with respect to the computational
basis it's often convenient to say that $\ket{x}$, instead of $x$, has been observed.
This abuse extends to the notation
for the probability that $m$ is observed.

We now consider how  single qubit systems may be put together to build larger systems.
\be[{\bf Postulate }\bf 1:]
\setcounter{enumi}{3}
\item {\bf Composite systems}\label{p:4} ~\\[.5em]
\hspace*{-.5in}\parbox{.9\textwidth}{
Given quantum mechanical systems associated to vector spaces $V$ and $W$ there is a composite quantum
mechanical system associated to $V\tens W$.
}
\ee
By induction this extends to composites of any finite number of systems. The composite of $2$ single
qubit systems is $\CC\ZZ_2\tens \CC\ZZ_2$, a $4$--dimensional space with basis vectors
$\ket{00}$,$\ket{01}$, $\ket{10}$,$\ket{11}$. We call this a $2${\em --qubit system}
or $2${\em --qubit quantum register}.
Similarly an $n${\em --qubit system} or {\em quantum register} is a copy of
the $2^n$--dimensional space $\CC\ZZ^{\tens n}_2$, which, as described
in Section \ref{subs:lin_alg}, has basis $\{\ket{i}:i=0,\ldots, 2^n-1\}$.
We shall by default assume that $n$--qubit system is equipped with this standard basis; and
immediately introduce an exception.
Given an $m$--qubit system $V_m$ and an $n$--qubit system $V_n$, we may form the
$mn$--qubit system $V_m\tens V_n$, which is naturally equipped with the basis
$\{\ket{ij}: i=0,\ldots ,2^m-1, j=0,\ldots ,2^n-1\}$.

An $n$--qubit system is the quantum analogue of the classical $n$--bit computer. Note that whereas
$n$--bits can contain, at any one time, one of $2^n$ possible values, a quantum computer can be
in a superposition of all of these values, in principle in
infinitely many different ways.
This completes our description of quantum mechanical systems.
We shall investigate some of the consequences
in the next few sections.
\subsection{Phase factors}\label{subs:phase}
Suppose that we have
measurement $\{M_m : m=1,\ldots,k\}$  of an $n$--qubit system.
The probability that $m$ is observed when the system is in state $\ket{x}$
is then
\begin{align*}
p(m,\ket{x})
& =\bket{x}{M_m^\dagger M_m|x}\\
&=\bket{e^{i\theta}x}{M_m^\dagger M_m|e^{i\theta}x}\\
&=p(m,e^{i\theta}\ket{x}),
\end{align*}
for any real number $\theta$. We call a complex number of modulus $1$ a
{\em phase factor}. Therefore, as far as the probabilities of
what is observed are concerned,
multiplication by a phase factor has no effect.

Also,
$T(e^{i\theta}\ket{x})=e^{i\theta}T\ket{x}$,
for any linear transformation $T$. Hence, if the system starts in state $\ket{x}$ and
after evolution or measurement (or both) its final state is $\ket{y}$ then
we can say that
starting in state $e^{i\theta}\ket{x}$ the final state is $e^{i\theta}\ket{y}$.

Therefore the introduction of the phase factor $e^{i\theta}$ is essentially
invisible to the quantum computer. Consequently we regard the states $\ket{x}$ and
$e^{i\theta}\ket{x}$ as the same state. We now see that after a measurement
in the computational basis, as in Section \ref{subs:lin_alg}, we may assume that
the quantum system projects to a basis vector of the system (with coefficient $1$).

Note that what we have said about phase factors
applies only to scalar multiples of the entire unit vector
which comprises the state. Altering individual coefficients of a state by a phase
factor may indeed change the state. For example $i(\quzero +\quone)/\sqrt{2}$ and
$(\quzero +\quone)/\sqrt{2}$ both correspond to the same state of a quantum system
but are not the same as $(\quzero +i\quone)/\sqrt{2}$.
\subsection{Multiple measurements}\label{subs:multi_measure}
We'll often have cause to employ more than one measurement as part of a single algorithm. One obvious
question is whether or not applying first one measurement and then a second is equivalent to applying
a single measurement. The answer is yes, and the single measurement is the obvious one,
as we shall now see. Let $M=\{M_m:m=1,\ldots, k\}$ and
$N=\{N_n:n=1,\ldots,l\}$ be measurements. Then, from the definition of measurement, we have
\[\sum_{m,n}\left(M_mN_n\right)^\dagger\left(M_mN_n\right)
=\sum_{m,n}N_n^\dagger M_m^\dagger M_mN_n=
\sum_n N_n^\dagger(\sum_m M_m^\dagger M_m)N_n=\sum_nN_n^\dagger I N_n=I.\]
Therefore $L=\{L_{(n,m)}=M_mN_n:m=1,\ldots,k, n=1,\ldots ,l\}$ is a measurement.

We claim that measurement using $N$ followed by measurement using $M$ is equivalent to
measurement using $L$. Suppose then that our system is in state $\ket{x}$ and that we first
measure using $N$. Then the probability of observing $n$ is
\[p_N(n)=\bket{x}{N^\dagger_nN_n| x}\]
and, if $n$ is observed, the system will then be in state
\[\ket{y}=\frac{N_n\ket{x}}{\sqrt{p_N(n)}}.\]
Now, with the system in state $\ket{y}$,
making a measurement using $M$ the probability of observing $m$ is
\[p_M(m)= \bket{y}{M^\dagger_mM_m| y}=\frac{1}{p_N(n)}\bket{x}{N_n^\dagger M_m^\dagger M_mN_n|x}\]
and, if $m$ is observed, the system will be in state
\[\ket{z}=\frac{M_m\ket{y}}{\sqrt{p_M(m)}}=\frac{M_mN_n\ket{x}}{\sqrt{p_M(m)}\sqrt{p_N(n)}}.\]
Hence the probability of measuring $n$ and then $m$, or $(n,m)$, is
\[p_N(n)p_M(m)= \bket{x}{N_n^\dagger M_m^\dagger M_mN_n|x}\]
which is, by definition, the probability $p_L(n,m)$ of observing $(n,m)$ using $L$. Also, if
$(n,m)$ is observed using $L$ then the system will be in state
%
\[\frac{M_mN_n\ket{x}}{\sqrt{p_L(n,m)}}=\ket{z}.\]
Therefore measurement using $L$ is equivalent to measurement using $N$ then $M$.

\subsection{Distinguishing states}
In classical computation it is reasonable to assume that if we have a register which may
take one of two distinct values then we can tell which of these values the register contains.
One consequence of the measurement postulate is that this is not always the case in quantum
computation. In fact we can tell apart orthogonal states but we cannot tell apart states which are
not orthogonal.

Following \cite{nielsenchuang}, to see that we can't distinguish non--orthogonal states,
assume that we have a quantum system which we know contains either the state $\ket{\psi_1}$ or
the state $\ket{\psi_2}$, and that we know what both of these states are. If $\ket{\psi_1}$ and
$\ket{\psi_2}$ are not orthogonal then we can write $\ket{\psi_2}=\a\ket{\psi_1}+\b\ket{\theta}$,
for some unit vector $\ket{\theta}$ orthonormal to $\ket{\psi_1}$
and $\a,\b\in\CC$, with $\a\neq 0$ and so $|\b|< 1$. Now assume that
we have a measurement $\{M_m:m=1,\ldots,k\}$ with which we can distinguish between $\ket{\psi_1}$ and
$\ket{\psi_2}$. That is to say we have sets $S_1$ and
$S_2$ such that $\{1,\ldots ,k\}=S_1\cup S_2$, where $S_1\cap S_2=\emptyset$ and
we observe $m\in S_i$ if and only if the system is in state $\ket{\psi_i}$ (before the measurement).

If we define $E_i=\sum_{m\in S_i} M^\dagger_mM_m$, for $i=1,2$, then
\begin{equation}\label{e:nodisting}
\bket{\psi_i|E_i}{\psi_i}=1, \,\textrm{ for } \, i=1,2,
\end{equation}
since the probability of observing $m\in S_i$ is $1$ if the system is in state
$\ket{\psi_i}$.
Also, $E_1+E_2=I$ so $\bket{\psi_i}{E_1+E_2|\psi_i}=1$,  for $i=1,2$, which implies that
\[\bket{\psi_1|E_2}{\psi_1}=0=\bket{\psi_2|E_1}{\psi_2}.\]
As $M_m^\dagger M_m$ is a positive operator (see \cite{kadisonringrose}, for example) so
is $E_i$ and so $\sqrt{E_i}$ is defined (and is a self--adjoint).
We now have
\[\bket{\psi_1|E_2}{\psi_1}=\bket{\psi_1}{\sqrt{E_2}^\dagger\sqrt{E_2}|\psi_1}=0,\]
so $\sqrt{E_2}\ket{\psi_1}=0$.
This implies that $\sqrt{E_2}\ket{\psi_2}=\sqrt{E_2}(\a\ket{\psi_1}+\b\ket{\theta})=\b\sqrt{E_2}\ket{\theta}$ so
\[\bket{\psi_2}{E_2|\psi_2}=|\b|^2\bket{\theta}{E_2|\theta}\le |\b|^2<1,\]
contrary to \eqref{e:nodisting} (where the first inequality follows from
\eqref{e:measurement}.)

The conclusion is that there is no such measurement. On the other hand, suppose we start with known states
$\ket{\psi_1}, \ldots, \ket{\psi_k}$ which are pairwise orthogonal. To see that we can distinguish between them
define a  measurement $\{M_m:m=0,\ldots,k\}$, where $M_m=\kbra{\psi_i}{\psi_i}$, $m\ge 1$, and
$M_0=I-\sum_{m=1}^{k}M_m$. Then, if the system is in state $\ket{\psi_i}$ the probability of measuring
$i$ is $\bket{\psi_i|M^\dagger_iM_i}{\psi_i}=1$. Thus, given that we know the system is in one
of these $k$ states, this measurement allows us to determine which one.

\subsection{No cloning}
Many classical algorithms make use of a copy function which replicates, or clones,
the contents of a given register, whatever they happen to be. This can be
achieved using the classical conditional not gate, CNOT, as follows.
Suppose
we have a $1$--bit register, that is a copy of $\ZZ_2$, and we wish to copy
its contents. Define the
function   CNOT from $\ZZ_2\bigoplus\ZZ_2$ to itself by CNOT$(x,y)=(x,x\oplus y)$, where
$\oplus$ denotes addition modulo $2$. To our $1$--bit register we adjoin a second
 $1$--bit register in state $0$: so if the first register is in state $x$ we now
have a $2$-bit register, that is $\ZZ_2\bigoplus\ZZ_2$, in state $(x,0)$. Applying
the conditional not function to this register we obtain
\[\textrm{CNOT}(x,0)=(x,x).\]
Both the first and  second $1$--bit registers now contain the original contents
of the first register. This process easily generalises to allow copying of $n$--bit
registers.

The question is whether we can do the same thing on a quantum computer. This would mean,
given an $n$--qubit register $V$, finding some fixed state $\ket{s}$ of $V$ and a
unitary transformation $U$ of $V\tens V$ such that
\begin{equation}\label{e:clone}
U\ket{x}\ket{s}=\ket{x}\ket{x}, \,\textrm{ for all }\, x\in V.
\end{equation}
It turns out that, as  a consequence of the following lemma, this is impossible.
\begin{lemma}
Let $\ket{x_1}$ and $\ket{x_2}$  be distinct unit vectors in $V$ and let $T$ be a
unitary transformation of $V\tens V$. If there is a unit vector
$\ket{s}\in V$ such that $T\ket{x_i}\ket{s}=\ket{x_i}\ket{x_i}$, for $i=1$ and $2$, then
$\ket{x_1}$ and $\ket{x_2}$ are orthogonal.
 \end{lemma}
\begin{pf}
Since $T$ is unitary we have
\[\bket{x_1 s}{x_2 s}
=\bket{ x_1 s}{T^\dagger T|x_2 s}
=\bket{x_1 x_1}{ x_2 x_2}.\]
That is, using \eqref{e:inner_tens},
\[\bket{x_1}{x_2}=\bket{x_1}{x_2}\bket{s}{s}=(\bket{x_1}{x_2})^2,\]
so $\bket{x_1}{x_2}=0$ or $1$. As $\ket{x_1}\neq \ket{x_2}$, and $\ket{x_1}$ and
$\ket{x_2}$ are unit vectors, it follows from the Cauchy--Schwarz inequality that
$\ket{x_1}$ and $\ket{x_2}$ are orthogonal, as required.
\end{pf}

Now let $U$ be a unitary transformation of $V\tens V$ and,
for a fixed $\ket{s}\in V$, set
$R=\{\ket{x}\in V\,:\, U\ket{x}\ket{s}=\ket{x}\ket{x}\}$.
The dimension of $V$ is $2^n$ so it follows, from the lemma above,
that $|R|\le 2^n$. This applies to any unitary transformation $U$ and any $\ket{s}\in V$,
so there can be no $U$ satisfying \eqref{e:clone}. Moreover we can copy at most
$2^n$ fixed, predetermined states, but they must be orthonormal. In fact, given
a set $\{ \ket{x_i}: i=1,\ldots ,2^n\}$ of $2^n$ orthonormal vectors of $V$,
we have a set $\{ \ket{x_i}\ket{s}: i=1,\ldots ,2^n\}$ of $2^n$ orthonormal vectors
of $V\tens V$. We can
extend this set to a basis of $V\tens V$.
Since $\{\ket{x_i}\ket{x_i}:i=1,\ldots, 2^n\}$ can also be extended to a
basis of $V\tens V$, we may
define a unitary transformation
of $V\tens V$ which maps $\ket{x_i}\ket{s}$ to $\ket{x_i}\ket{x_i}$, for all $i$.
Hence we may copy up to $2^n$ fixed orthonormal states.

In a classical probabilistic algorithm we may repeat a calculation a fixed number
of times
(in order to have a high probability of success) by first copying the input and
then performing the calculation on each copy. However on a quantum computer,
due to the fact that we cannot copy arbitrary states, to repeat the calculation
we must repeat the entire algorithm.

\subsection{Entangled states}

Let $x\in V\tens W$. Then $x$ is said to be {\em disentangled} if there exist
$v\in V$ and $w\in W$ such that $x=v\tens w$. Otherwise $x$ is said to be
{\em entangled}. The definition extends in the
obvious way to $n$-fold tensor products. An
entangled unit vector of
an $n$--qubit system is
called
an {\em entangled state}.
For example, $|00\rangle+|11\rangle$ is an entangled state. In
fact, if there exist $a,b,c,d\in \CC$ such that
\begin{eqnarray*}
|00\rangle+|11\rangle & = &
(a\quzero+b\quone)\tens(c\quzero+d\quone) \\
& = & ac|00\rangle+bc|10\rangle+ad|01\rangle+bd|11\rangle,
\end{eqnarray*}
then $bc=ad=0$ and $ac=bd=1$, which is impossible.

Entangled states play an important role in quantum teleportation and in
super--dense coding. For details of these applications see, for example,
\cite{nielsenchuang} or \cite{braunsteinetal}.

\subsection{Observing multiple qubit systems}

A quantum algorithm may use different parts of its
register for different purposes and so it is often convenient to view a quantum
system as a composite of sub--systems. This amounts to decomposing
the vector space comprising the system into a tensor product
of vector sub--spaces. The purpose of doing this is usually so that measurement
may be taken on one part of the system but not the other. We now consider how
this may be arranged.

Suppose that the $mn$--qubit system $V$ is the tensor product of $m$ and $n$--qubit systems
$Q$ and $R$, respectively. Assume that we
have measurements $M=\{M_a:a=1,\ldots,k\}$ and $N=\{N_b:b=1,\ldots, l\}$ of
$Q$ and $R$. Then we may define measurements
$\hat M=\{\hat M_a=M_a\tens I:a=1,\ldots,k\}$
and $\hat N=\{\hat N_b=I\tens N_b:b=1,\ldots, l\}$ of $Q\tens R$. As in Section
\ref{subs:multi_measure} we obtain a measurement $L_{ab}=\{\hat M_a \hat N_b
:a=1,\ldots, k, b=1,\ldots l\}$. This time
we have
$\hat M_a\hat N_b=(M_a\tens I)(I\tens N_b)=(I\tens N_b)(M_a\tens I)=\hat N_b\hat M_a$, for
all $a,b$. Therefore, as in Section \ref{subs:multi_measure} measurement with
$\hat N$ followed
by
$\hat M$ is equivalent to measurement with $L$ and, in addition, the same is true of
measurement with $\hat M$ followed by $\hat N$. We call a measurement $\hat M$
induced in this way from a measurement on $Q$ {\em a measurement of register}
$Q$ and say that $a$ is {\em observed in register} $Q$.
\begin{ex}
Suppose that we have a $2$--qubit system $V=\CC\ZZ_2\tens \CC\ZZ_2$ and
$Q=R=\CC\ZZ_2$ and we have measurements $M=\{M_0,M_1\}$ and $N=\{N_0,N_1\}$, where
$M_i=N_i=\kbra{i}{i}$. Then $\hat M_i=\kbra{i}{i}\tens I$,
$\hat N_j=I\tens \kbra{j}{j}$ and $L_{ij}=\kbra{i}{i}\tens\kbra{j}{j}$.
Consider the state
\[\ket{\psi}= a|00\rangle+b|01\rangle+c|10\rangle+d|11\rangle. \]
where $|a|^2+|b|^2+|c|^2+|d|^2=1$.
To see what happens if we measure with $\hat M$ we may rewrite $\ket{\psi}$ as
\[ \quzero\tens(a\quzero+b\quone)+
  \quone\tens(c\quzero+d\quone).\]
Then it's clear that $\hat M_0\ket{\psi}=\quzero\tens(a\quzero+b\quone) $ and
$\hat M_1\ket{\psi}=\quone\tens(c\quzero+d\quone).$ Hence
\[p_{\hat M}(\ket{\psi}\ra 0)=|a|^2+|b|^2=u^2\mbox{ and }
   p_{\hat M}(\ket{\psi}\ra 1)=|c|^2+|d|^2=v^2, \]
where $u,v\ge 0$.

Observations of $0$ or $1$ in register $Q$ will therefore result in
the quantum state
\[ \quzero\tens\left(\frac{a}{u}\quzero+\frac{b}{u}\quone\right)\mbox{ or }
   \quone\tens\left(\frac{c}{v}\quzero+\frac{d}{v}\quone\right)
   \mbox{ respectively.}
\]

Similarly, if $w,x\ge 0$ are such that $w^2=|a|^2+|c|^2$ and $x^2=|b|^2+|d|^2$
then measuring $V$ with $\hat N$ the probabilities of observing $0$ and $1$
in register $R$ are
\[p_{\hat N}(\ket{\psi}\ra 0) =w^2\mbox{ and }
   p_{\hat N}(\ket{\psi}\ra 1)=x^2. \]
If $0$ or $1$ is observed in register $R$ then the resulting  quantum state is
\[ \left(\frac{a}{w}\quzero+\frac{c}{w}\quone\right)\tens\quzero
   \mbox{ or }
   \left(\frac{b}{x}\quzero+\frac{d}{x}\quone\right)\tens\quone
   \mbox{ respectively.}
\]
\end{ex}

We now consider the effect of measurements in registers $Q$ and $R$
on a disentangled state $\ket{xy}=\ket{x}
\tens \ket{y}$, where $\ket{x}\in Q$ and $\ket{y}\in R$.
We shall see that in this case the probabilities of observing $a$ in register
$Q$ or $b$ in register $R$ are independent.
 We
have $\hat M_a^\dagger \hat M_a=M_a^\dagger M_a\tens I$ so the probability
of observing $a$ if we measure $V$ with $\hat M$ is
\[p_{\hat M}(\ket{xy}\ra a)=\bket{xy}{\hat M_a^\dagger \hat M_a|xy}=
\bket{x\tens y}{M_a^\dagger M_a\tens I|x\tens y}=
\bket{x}{M_a^\dagger M_a|x}\bket{y}{y}.\]
Since $\ket{xy}$ is a unit vector we may assume that $\ket{x}$ and $\ket{y}$ are
unit vectors, by writing
\[\ket{xy}=\frac{1}{|\ket{xy}|}\ket{xy}=\frac{\ket{x}}{|\ket{x}|}
\frac{\ket{y}}{|\ket{y}|},\]
if necessary. Hence $\bket{y}{y}=1$ and
\[p_{\hat M}(\ket{xy}\ra a)=\bket{x}{M_a^\dagger M_a|x}=p_M(\ket{x}\ra a),\]
the probability that $a$ is observed if $Q$ is measured with $M$.
Similarly
\[p_{\hat N}(\ket{xy}\ra b)=p_N(\ket{y}\ra b)\]
the probability that $b$ is observed if $R$ is measured with $N$.
The probability of observing both $a$ in register $Q$ and $b$ in register
$R$ when $V$ is in state $\ket{xy}$ is
\begin{align*}
p_L(ab)=\bket{xy}{L_{ab}^\dagger L_{ab}|xy}
& =\bket{x\tens y}{M_a^\dagger M_a\tens N^\dagger_bN_b|x\tens y}\\
&=
\bket{x}{M_a^\dagger M_a}\bket{y}{N_b^\dagger N_b|y}\\
&=p_{\hat M}(a)p_{\hat N}(b).
\end{align*}
Hence $p_{\hat M}(\ket{xy}\ra a)$ and $p_{\hat N}(\ket{xy}\ra b)$ are independent.

\begin{ex} With $V$, $Q$, $R$, $M$ and $N$ as in the previous example,
let $\ket{x}=\a\quzero +\b\quone$ and $\ket{y}=
\c\quzero +\d\quone$, where
where $|\a|^2+|\b|^2=1$ and $|\c|^2+|\d|^2=1$.
Then
\[ \ket{x}\tens \ket{y}=\alpha\gamma|00\rangle+\alpha\delta|01\rangle
    +\beta\gamma|10\rangle+\beta\delta|11\rangle.\]
Hence
\[p_{\hat M}(0)=|(\a\c)|^2+|\a\d|^2 \, \textrm{ and }\,
p_{\hat N}(0)=|(\a\c)|^2+|\b\c|^2,\]
while
\[p_L(00)=|\a\c|^2.\]
Hence
\[p_{\hat M}(0)p_{\hat N}(0)=|\a\c|^2(|\a\c|^2+|\b\c|^2 +|\a\d|^2+|\b\d|^2)
=|\a\c|^2=p_L(00).\]
\end{ex}

On the other hand, in general,  the measurements of individual registers
do not have independent probabilities. To see this  consider the entangled
state  $(\ket{00}+\ket{11})/\sqrt{2}$ in the system of the previous example.
Measuring in register $Q$ we observe either $0$ or $1$, each with a probability
of $1/2$. The same holds for register $R$. If these probabilities are independent
then the probability of observing $00$ should be $1/4$.
However, it's easy to see that the probability of observing
$00$ is also $1/2$. In fact after the first measurement the system projects to
one of the states $\ket{00}$ or $\ket{11}$. Measuring the former we observe $00$
with probability $1$ and $11$ with probability $0$.

The above example illustrates the fact that if $\ket{\psi}$ is
disentangled then probabilities of observing $a$ in  register $Q$
 and $b$ in register $R$ are independent, where $\ket{a}$ and $\ket{b}$
range over basis vectors of $Q$ and $R$. The converse is {\em not} true.
For example take $\ket{\psi}=[\ket{00}+\ket{01}+\ket{10}-\ket{11}]/2$.
The probability $p_{\hat M}(a)$
of observing $a$ in register $Q$ is $1/2$, as is the probabilty $p_{\hat N}(b)$
of observing $b$ in register $R$, for all $a$ and $b$ in $\{0,1\}$.
As the probabilty $p_{L}(ab)$ of observing $\ket{ab}$ is $1/4$, the
probabilities $p_{\hat M}$ and $p_{\hat N}$ are independent. However
$\ket{\psi}$ is entangled.

\subsection{Quantum parallelism}\label{subs:parallel}
We develop here the notion of parallel quantum computation which we touched on briefly in
Section \ref{subs:over}. A basic tool in the operation of this scheme is the family of functions
we describe next.

The
{\em single qubit Walsh--Hadamard transformation} is the unitary operator $W$ on a single qubit system
given by
\begin{align}
\nonumber
W(\quzero) &=\frac{1}{\sqrt{2}}\left(\quzero+\quone\right)\mbox{ and }\\
   W(\quone) & =\frac{1}{\sqrt{2}}\left(\quzero-\quone\right). \label{e:W_H_1}
\end{align}

We can express $W$ more concisely by writing
\begin{align*}
W(\ket{x}) & = \frac{1}{\sqrt{2}}\left(\quzero+(-1)^x\quone\right)\\
&=\frac{1}{\sqrt{2}}\sum_{k=0}^1(-1)^{kx}\ket{k}.
\end{align*}
It is easy to see by direct calculation that $W$ is an involution, that is $W^2=I_2$.
Moreover, if we ignore the complex coefficients $W$ is reflection in the line which
makes an angle of $\pi/8$ with the $\quzero$--axis.

The {\em n--bit Walsh--Hadamard transformation} $W_n$ is defined to be $W^{\tens n}$.
As $W$ is an involution we have $W_n^2=I_2^{\tens n}=I_{2^n}$, so $W_n$ is also an involution.
Applied to $\quzero^{\tens n}$, $W_n$ generates a uniform linear combination of the
integers from $0$ to $2^n-1$, i.e.
\[ W_n(|0\cdots 0\rangle)=\frac{1}{\sqrt{2^n}}\sum_{x=1}^{2^n-1}|x\rangle. \]
For example,
\begin{eqnarray*}
W_2|00\rangle & = & (W\tens W)(\quzero\tens\quzero) \\
& = & W\quzero\tens W\quzero \\
& = & \frac{1}{2}\left( (\quzero+\quone)\tens(\quzero+\quone) \right) \\
& = & \frac{1}{2}\left( \quzero\tens\quzero+\quzero\tens\quone+\quone\tens\quzero+\quone\tens\quone\right) \\
& = & \frac{1}{\sqrt{2^2}}\left( |00\rangle+|01\rangle+|10\rangle+|11\rangle \right) .
\end{eqnarray*}
This generalises in the obvious way to $W_n$ and allows us, starting with the simple basis
state $\ket{0\cdots 0}$, to prepare a uniform superposition of all basis vectors.

For computation a more concise notation is convenient and to this end we define the
following notation. Let $\ket{x}$ and $\ket{y}$ be basis vectors in an $n$--qubit system, where
$x$ and $y$ are $n$--bit binary integers. Define
\[x\cdot y=\sum_{i=0}^{n-1} x_iy_i.\]
(Note that this is not the inner product of $\ket{x}$ and $\ket{y}$
(see page \pageref{e:tinnerp}).
It extends to a symmetric bilinear form on $\ket{x}$ and $\ket{y}$ regarded as vectors in
a $2^n$--dimensional
space over $\ZZ_2$. However it may be zero when $\ket{x}$ and $\ket{y}$ are both non--zero.)
Now, setting $m=2^n$, we have
\begin{align}
W_n\ket{x} &=\btens_{i=0}^{n-1}W\ket{x_i} \nonumber\\
&=  \btens_{i=0}^{n-1}\frac{1}{\sqrt{2}}\sum_{k_i=0}^1(-1)^{k_ix_i}\ket{k_i}\nonumber\\
&= \frac{1}{\sqrt{2^n}}\sum_{k_0=0}^1\cdots \sum_{k_{n-1}=0}^1 (-1)^{x_0k_0}\cdots (-1)^{x_{n-1}k_{n-1}}\ket{k_0\cdots k_{n-1}}\nonumber\\
&= \frac{1}{\sqrt{2^n}}\sum_{k_0\cdots k_{n-1}=0}^{m-1}(-1)^{x_0k_0}\cdots (-1)^{x_{n-1}k_{n-1}}\ket{k_0\cdots k_{n-1}}\nonumber\\
&= \frac{1}{\sqrt{2^n}}\sum_{k=0}^{m-1} (-1)^{x\cdot k}\ket{k}.\label{e:W_H_n}
\end{align}
%

The Walsh--Hadamard functions allow us to prepare the input to parallel computations. Now we consider
the computations themselves.
Let $f:\ZZ_2^m\ra\ZZ_2^k$ be a function, not necessarily invertible. As we're not assuming that
$f$ is invertible we cannot use it, as it is, as a transformation in a quantum computer. However, at the
expense of introducing some extra storage space we can devise a unitary transformation to simulate
$f$.
We require a quantum system $V$ which is the tensor product of an
$m$--qubit quantum system with a
$k$--qubit quantum system. 
Recall that $V$ has basis consisting of the vectors $\ket{x}\tens \ket{y}$,  where
$x$ and $y$ are binary representations of integers in $\{0,\ldots,2^m-1\}=\ZZ_2^m$
and $\{0,\ldots,2^k-1\}=\ZZ_2^k$ respectively.
Define the linear
transformation
\[
U_f:\ket{x}\tens\ket{y}\mapsto \ket{x}\tens\ket{y\oplus f(x)},
\]
where $\oplus$ denotes addition in the group $\ZZ_2^k$ (known as
``bitwise exclusive OR'' in the literature).
For fixed $x$, we see that $y\oplus f(x)$ takes every value in $\ZZ_2^k$ exactly once,
 as $y$ varies over $\set{0,\ldots,2^k-1}$.
Therefore $U_f$ simply permutes all $2^{m+k}$ basis elements of $V$ and it follows that it is unitary.
Moreover $U_f(\ket{x}\tens \ket{0})=\ket{x}\tens \ket{f(x)}$ and in this sense $U_f$ simulates
$f$.
The map $U_f$ is referred to as the {\em standard oracle} for the function $f$.
The standard oracle may thus be used to simulate
any function, invertible or not, on a quantum computer. It follows that
any function which may be carried out by a classical computer may also be carried out by a
quantum computer.

In the case where $f$ is a bijection, and only in this case, we
can define the simpler and more obvious oracle
$\ket{x}\ra\ket{f(x)}$. This is called the {\em minimal} or {\em erasing oracle} for $f$.
Its relation to the standard
oracle is considered in \cite{kashefietal}. Furthermore, in \cite{aaronson} a problem is given in which a minimal oracle is shown to be
exponentially more powerful than a standard oracle.) The form of $U_f$ above may seem strange, but in fact it originates in classical
reversible computing and has been adapted for the purposes of quantum computing. See \cite{kitaev} for more details of reversible
computing.

If we apply $U_f$ to
\[ W_m(\quzero^{\tens m})\tens \quzero^{\tens k}\]
we obtain
\begin{eqnarray*}
U_f\left(\frac{1}{\sqrt{2^m}}\sum_{x=0}^{2^m-1}\ket{x}\tens\ket{0}\right) & = &
\frac{1}{\sqrt{2^m}}\sum_{x=0}^{2^m-1}U_f(\ket{x}\tens\ket{0}) \\ & = &
\frac{1}{\sqrt{2^m}}\sum_{x=0}^{2^m-1}\ket{x}\tens\ket{f(x)}.
\end{eqnarray*}
We can view this as a simultaneous computation of $f$ on all possible values of $x$,
although the fact that
$\ket{f(x)}$ is associated with the state $\ket{x}$, for all $x$, may sometimes be a problem.
Creation of this kind of state is often referred to as {\em quantum parallelism} and is an easy and
standard first step in many quantum computations. The tricky part is to glean useful information from this
(extremely entangled) output state.
\begin{ex}
Suppose that $m=k=2$, so that $\ZZ_2^m=\ZZ_2^k=\ZZ_2^2=\set{0,\ldots,3}$. Let $f$ be the
function defined by $f(0)=1$, $f(1)=2$, $f(2)=0$ and $f(3)=3$. The  quantum system $V$ is
the tensor product $\CC\ZZ_2^2\tens\CC\ZZ_2^2$ and has basis $\{\ket{x}\tens\ket{y}:0\le x,y\le 3\}$.
We can represent the elements of this basis on a $3\times 3$ grid, with $x$ indexing the horizontal squares and
$y$ the vertical ones. For many of the algorithms we consider, the quantum
state is always in a uniform superposition of an $r$--element subset of the set of
basis elements, with phase (coefficient) $\pm 1/\sqrt{r}$.
We represent such states by using a black square
for a coefficient of $1/\sqrt{r}$, a grey square for a coefficient of $-1/\sqrt{r}$
and a white square for a coefficient of $0$. For example,
a basis state is represented by a single square as in Figure~\ref{fig:basisstates} and
the state $(W\ket{0})\tens\ket{0}$ is represented as in Figure~\ref{fig:uniformx}.
If we apply $U_f$ to this state we obtain the state shown in Figure~\ref{fig:graphstate},
which can be considered as a uniform superposition over
all of the points of the graph of $f$.

\begin{figure}
\begin{center}
\psfrag{x}{$x$}
\psfrag{y}{$y$}
\psfrag{s1}{$\quzero\tens\quone$}
\psfrag{s2}{$-\ket{3}\tens\ket{2}$}
\includegraphics[width=2.1in]{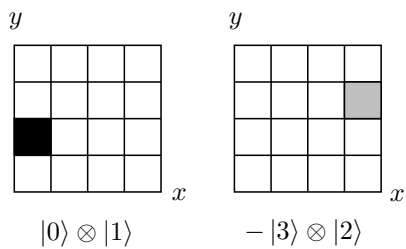}
\caption{\label{fig:basisstates} Basis states.}
\end{center}
\end{figure}
\begin{figure}
\begin{center}
\psfrag{x}{$x$}
\psfrag{y}{$y$}
\psfrag{s1}{$(W\quzero)\tens\quzero$}
\includegraphics[width=1in]{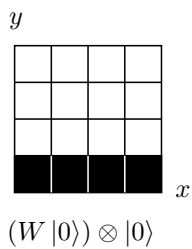}
\caption{\label{fig:uniformx} A uniform superposition in the first register}
\end{center}
\end{figure}
\begin{figure}
\begin{center}
\psfrag{x}{$x$}
\psfrag{y}{$y$}
\psfrag{s1}{$U_f((W\quzero)\tens\quzero)$}
\includegraphics[width=1.2in]{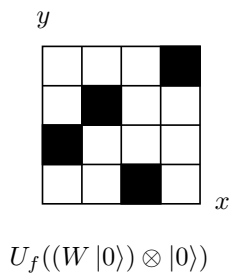}
\caption{\label{fig:graphstate} A quantum state corresponding to the graph of $f$}
\end{center}
\end{figure}
It's useful, as an exercise, to calculate $U_f$ applied to $(W\ket{0})\tens\ket{1}$, $(W\ket{0})\tens\ket{2}$ and $(W\ket{0})\tens\ket{3}$
in turn, to understand how $U_f$ is constructed, and why it is reversible. In fact, $U_f$ is always an involution
(i.e. $U_f\circ U_f=I$),
regardless of what $f$ is.
\end{ex}

\begin{ex} Pictures of the quantum system, as in the previous example, can also be used to help understand measurements of
quantum states, and how they relate to entanglement.
All measurements are taken with respect to the computational basis (see Section \ref{subs:postulates}).
A disentangled state such as, for example,
\[\ket{2}\tens\frac{1}{\sqrt{3}}\left(\ket{0}+\ket{1}+\ket{3}\right)\] looks the same in every nonzero column (up to phase), so
measuring the first register (which always results in ${2}$) doesn't affect the distribution of
non--zero coefficients in the second register
(see Figure~\ref{fig:measpic1}).
\begin{figure}
\begin{center}
\includegraphics[width=1.0in]{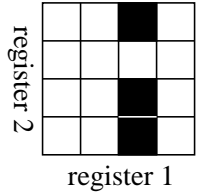}
\caption{\label{fig:measpic1} A disentangled state}
\end{center}
\end{figure}
On the other hand, if we have an entangled state such as, for example,
\[ \frac{1}{\sqrt 6}\left(\ket{0}\tens\ket{0}+\ket{0}\tens\ket{2}-\ket{2}\tens\ket{1}+\ket{2}\tens\ket{2}-\ket{2}\tens\ket{3}+
\ket{3}\tens\ket{3}\right) \]
we can see directly from Figure~\ref{fig:measpic2} how measurements of the first register affect the distribution on the second.
\begin{figure}[!btph]
\begin{center}
\psfrag{half}{{\small $\frac{1}{2}$}}
\psfrag{third}{{\small $\frac{1}{3}$}}
\psfrag{sixth}{{\small $\frac{1}{6}$}}
\psfrag{1}{{\small $1$}}
\psfrag{0}{{\small $0$}}
\psfrag{register1register1}{{Register 1}}
\psfrag{kx}{Probability of observing ${x}$}
\includegraphics[width=3in]{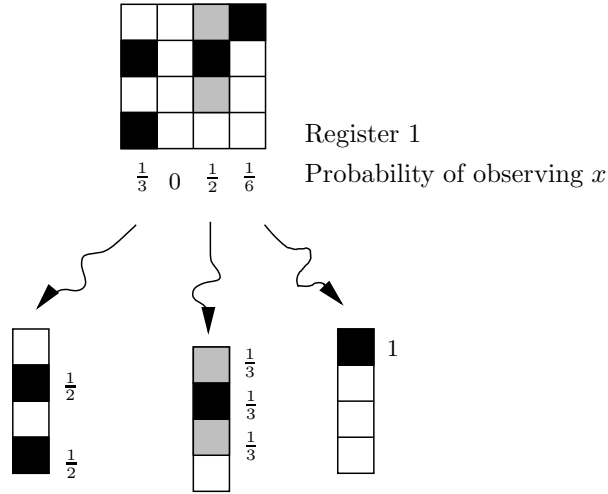}
\caption{\label{fig:measpic2} Measurement of an entangled state}
\end{center}
\end{figure}
If we observe ${0}$ in the first register, then there is a probability of ${1}/{2}$ of observing ${0}$
in the second register, and a probability of ${1}/{2}$ of observing ${2}$. If we observe ${2}$ in the first register
then we have probabilities of ${1}/{3}$ of observing either ${2}$, ${3}$ or ${4}$ and if we observe ${3}$ in
the first register then we observe ${3}$ with certainty in the second register.
\end{ex}

Some care is required in the interpretation of these diagrams. For example, the diagram of
the entangled state $\ket{00}-\ket{01}-\ket{10}+\ket{11}$ will suggest different results in the second
register after measurement
of the first register: depending on whether $0$ or $1$ is observed. However these results differ only by
a phase factor (of $-1$) so are, in fact, the same.
\newpage
\section{The Deutsch--Jozsa algorithm}

\subsection{Oracles and query complexity}
Deutsch \cite{deutsch} was the first to show that a speed--up in complexity is
possible when passing from classical to quantum computations. It is important to
understand that the complexity referred to is {\em query complexity}. The idea is
that we have a ``black box'' or ``oracle'', as described in Section \ref{subs:parallel},
which evaluates some function (classically this would be a function on integers,
but in the quantum scenario we allow evaluations on complex vectors). Query complexity simply
addresses how many times we have to ask the oracle to evaluate the function on some input, in order to determine some
property of $f$. It
ignores how many quantum gates we require to actually implement the function. For a genuine upper bound on time complexity
we must demonstrate efficient implementation of the oracle.
For example, this is the case in Shor's algorithm in the next section where we show that the quantum Fourier transform
can be implemented using a number of gates logarithmic in the size of the input.

Again, a lower bound on the query complexity of  a given algorithm, with respect to a specific oracle,
does not necessarily give a lower bound on time complexity of the algorithm. This  is because
we know nothing about the operation of the oracle: if we
knew how the oracle worked then perhaps we could see how to do without it.
Within the context of query complexity (relative to a specific oracle) many quantum algorithms have been
{\em proved} to be more efficient than any classical counterpart.
However, so far, not a single instance exists where we can say the same about true
time complexity. To do so would usually require a lower bound on the classical
complexity of a given problem and this often brings us up against
difficult open problems in classical complexity theory. For example, Shor's
algorithm is a (non--deterministic) quantum algorithm for factoring integers.
Although no classical polynomial time algorithm is {\em known}
 for this problem, whether or not such an algorithm exists is an open question.

Given two functions  $f$ and  $g$ from $\NN$ to $\RR$ we say that $f=O(g)$ if there
exist constants $c,k\in\RR$ such that $|f(n)|\le c|g(n)|+k$, for all $n\in\NN$.
We also say that $f=\Omega(g)$ if $g=O(f)$. Thus $O$ is used to describe upper
bounds and $\Omega$ to describe lower bounds.

\subsection{The single qubit case of Deutsch's algorithm}
\begin{figure}[!btph]
\begin{center}
\psfrag{x}{$x$}
\psfrag{Ux}{$Ux$}
\psfrag{U}{$U$}
\includegraphics[width=2.2in]{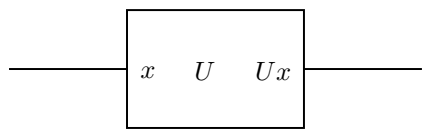}
\caption{\label{fig:singqbit} A single--qubit gate}
\end{center}
\end{figure}

\begin{figure}[!btph]
\begin{center}
\psfrag{mq}{$m$ qubits}
\psfrag{kq}{$k$ qubits}
\psfrag{k0}{$\ket{0}$}
\psfrag{x}{$x$}
\psfrag{y}{$y$}
\psfrag{yf}{$y\oplus f(x)$}
\psfrag{W}{$W$}
\psfrag{uf}{$U_f$}
\psfrag{aaaabbbbccccdddd}{$\displaystyle\frac{1}{\sqrt{2^m}}\displaystyle\sum_{x=0}^{2^m-1}\ket{x}\tens\ket{f(x)}$}
\includegraphics[width=3.6in]{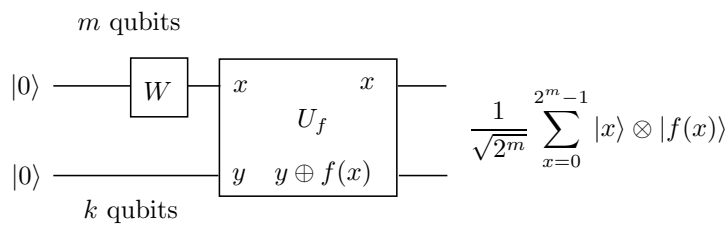}
\caption{\label{fig:qpar} The circuit for quantum parallelism}
\end{center}
\end{figure}

\begin{figure}[!btph]
\begin{center}
\psfrag{x}{$x$}
\psfrag{y}{$y$}
\psfrag{uf}{$U_f$}
\psfrag{mq}{$m$ qubits}
\psfrag{kq}{$k$ qubits}
\psfrag{k0}{$\ket{0}$}
\psfrag{yf}{$y\oplus f(x)$}
\psfrag{W}{$W$}
\includegraphics[width=2.8in]{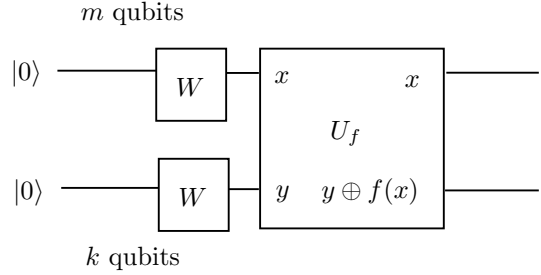}
\caption{\label{fig:noinfo} An insensitive quantum circuit}
\end{center}
\end{figure}

The unitary maps involved in quantum computing can often be
represented pictorially via {\em quantum circuit diagrams}.
An operator $U$ on a single
quantum register is represented as in Figure~\ref{fig:singqbit}.
We also draw gates for operators on two or more quantum registers:
the set--up for quantum parallelism as described in Section \ref{subs:parallel}
is shown in Figure~\ref{fig:qpar}.
A similar circuit is shown in
Figure~\ref{fig:noinfo}. Here an additional
Walsh--Hadamard gate operates on the second register, transforming
its contents into
a uniform superposition,  before entering the $U_f$ gate.

Computing the final state of this system we have
\begin{eqnarray*}
\ket{0}\tens\ket{0} & \stackrel{W\tens W}{\mapsto} & \frac{1}{\sqrt{2^{m+k}}}\sum_{x=0}^{2^m-1}\sum_{y=0}^{2^k-1}\ket{x}\tens\ket{y} \\
& \stackrel{U_f}{\mapsto} & \frac{1}{\sqrt{2^{m+k}}}\sum_{x=0}^{2^m-1}\sum_{y=0}^{2^k-1}\ket{x}\tens\ket{y\oplus f(x)} \\
& = & \frac{1}{\sqrt{2^{m+k}}}\sum_{x=0}^{2^m-1}\sum_{y=0}^{2^k-1}\ket{x}\tens\ket{y},
\end{eqnarray*}
because $y\oplus f(x)$ takes each possible value exactly once, as $y$ ranges over $\set{0,\ldots,2^k-1}$.
This computation can certainly not be used to gain any information about $f$,
because its final state is  the
same, whatever $f$ is.
 However, Deutsch showed that, with $k=1$, if  we alter input to the second register to $\ket{1}$
then we can obtain some information on the nature of $f$.

\begin{figure}[!btph]
\begin{center}
\psfrag{x}{$x$}
\psfrag{y}{$y$}
\psfrag{uf}{$U_f$}
\psfrag{1q}{$1$ qubit}
\psfrag{k1}{$\ket{1}$}
\psfrag{k0}{$\ket{0}$}
\psfrag{yf}{$y\oplus f(x)$}
\psfrag{W}{$W$}
\includegraphics[width=3in]{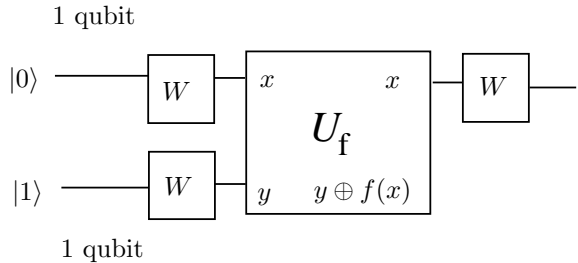}
\caption{\label{fig:deutsch} The quantum circuit for Deutsch's algorithm}
\end{center}
\end{figure}

Deutsch's algorithm concerns functions $f:\{0,1\}\ra\{0,1\}$. We call such a function
{\em constant}  if $f(0)=f(1)$ and {\em balanced} if $f(0)\not=f(1)$. Given such a function
suppose that we wish to
determine whether $f$ is constant or balanced (it must be one or the other). Classically, this
requires two evaluations of $f$. Let $U_f$ be the standard oracle
for $f$ (see Section \ref{subs:parallel}). We shall show that a quantum computer only
needs a single evaluation of the oracle to determine whether $f$ is constant or balanced (with certainty).
The quantum circuit for the algorithm is
shown in Figure~\ref{fig:deutsch}.
After passing through the Walsh--Hadamard gates, the registers are in the
state
\[ (W\tens W)(\quzero\tens\quone)=\left(\frac{\quzero+\quone}{\sqrt{2}}\right)
\tens\left(\frac{\quzero-\quone}{\sqrt{2}}\right). \]
Now, if $x\in\{0,1\}$, we have
\begin{eqnarray*}
U_f\left(|x\rangle\tens\left(\frac{\quzero-\quone}{\sqrt{2}}\right)\right)
& = & |x\rangle\tens\left(\frac{|0\oplus f(x)\rangle-
|1\oplus f(x)\rangle}{\sqrt{2}}\right) \\
& = & |x\rangle\tens\frac{1}{\sqrt{2}}\left(\left\{\begin{array}{ll}
\quzero-\quone & \mbox{if }f(x)=0 \\ \quone-\quzero &
\mbox{if }f(x)=1\end{array}\right.\right)\\
& = & (-1)^{f(x)}|x\rangle\tens\left(\frac{\quzero-\quone}{\sqrt{2}}\right)
\end{eqnarray*}
Therefore, by linearity, after passing through the $U_f$ gate the system is in state
\[
\left(\frac{(-1)^{f(0)}\quzero+(-1)^{f(1)}\quone}{\sqrt{2}}\right)\tens
\left(\frac{\quzero-\quone}{\sqrt{2}}\right)
\]
which is equal to
\[
\pm
\left(
  \frac{\quzero+\quone}{\sqrt{2}}
\right)
\tens
\left(
  \frac{\quzero-\quone}{\sqrt{2}}
\right)
,
\, \textrm{ if $f$ is constant,} \,
\]
and
\[
\pm
\left(
  \frac{\quzero-\quone}{\sqrt{2}}
\right)
\tens
\left(
  \frac{\quzero-\quone}{\sqrt{2}}
\right),
\, \textrm{ if $f$ is balanced.}
\]
After passing through the final Walsh--Hadamard gate, the
first qubit is in state
\[
\begin{array}{ll}
\pm\quzero & \mbox{if $f$ is constant} \\
\pm\quone & \mbox{if $f$ is balanced}
\end{array}
.
\]
So measuring this qubit, with respect to the computational basis, we observe $0$ with probability
$1$, if $f$ is constant,
and $1$ with probability $1$, if $f$ is balanced. Since $f$ has only been
evaluated once, this demonstrates that quantum computers are strictly more
efficient than classical computers, when we are referring to deterministic
black box query complexity.

\begin{figure}[!btph]
\begin{center}
\psfrag{f1}{$f_1$}
\psfrag{f2}{$f_2$}
\psfrag{f3}{$f_3$}
\psfrag{f4}{$f_4$}
\includegraphics[width=1.8in]{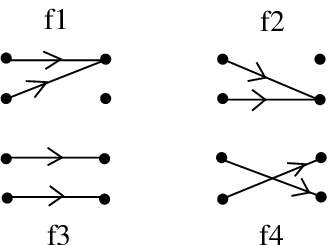}
\caption{\label{fig:deutschfs} The four possible functions $\set{0,1}\ra\set{0,1}$}
\end{center}
\end{figure}

\begin{figure}[!btph]
\begin{center}
\psfrag{constant}{constant}
\psfrag{balanced}{balanced}
\psfrag{lk0}{$\leadsto\ket{0}$}
\psfrag{lk1}{$\leadsto\ket{1}$}
\psfrag{x}{$x$}
\psfrag{y}{$y$}
\psfrag{u1}{$U_1$}
\psfrag{u2}{$U_2$}
\psfrag{u3}{$U_3$}
\psfrag{u4}{$U_4$}
\psfrag{IxW}{$I\tens W$}
\psfrag{WxI}{$W\tens I$}
\includegraphics[width=4.5in]{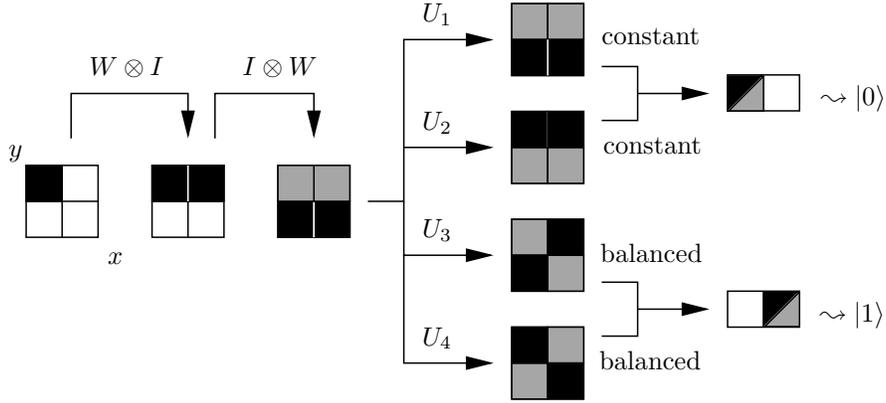}
\caption{\label{fig:deutsch2} Quantum states created by Deutsch's algorithm}
\end{center}
\end{figure}

In fact Deutsch's algorithm puts each of the functions in Figure~\ref{fig:deutschfs} in one of the two classes:
constant or balanced.
With the notation of Figure~\ref{fig:deutschfs}, write $U_i=U_{f_i}$.
Then the operation of Deutsch's algorithm on each function can be pictorially
represented as in Figure~\ref{fig:deutsch2}.

\subsection{The general Deutsch--Jozsa algorithm}\label{subs:gDJ}
\begin{figure}[!btph]
\begin{center}
\psfrag{x}{$x$}
\psfrag{y}{$y$}
\psfrag{uf}{$U_f$}
\psfrag{mq}{$m$ qubits}
\psfrag{1q}{$1$ qubit}
\psfrag{k1}{$\ket{1}$}
\psfrag{k0}{$\ket{0}$}
\psfrag{yf}{$y\oplus f(x)$}
\psfrag{W}{$W$}
\psfrag{Wm}{$W_m$}
\includegraphics[width=3in]{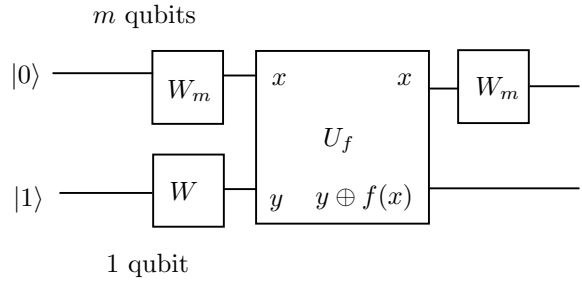}
\caption{\label{fig:deutschstate} The quantum circuit for the Deutsch--Jozsa algorithm}
\end{center}
\end{figure}

The generalisation of the algorithm of the previous section to $m$ qubits
is due to Deutsch and Jozsa \cite{deutschjosza} (see also \cite{cleveetal}
for the improved version which we present here). A function $f:\CC\ZZ_2^m\ra\{0,1\}$ is called
{\em balanced} if $|f^{-1}(0)|=|f^{-1}(1)|=2^{m-1}$.  Assume we know only that $f$ is either constant or
balanced, and we that wish to determine which of these properties $f$ has.
Classically this requires $2^{m-1}+1$ evaluations. However on  a quantum computer it
can be done with  a single evaluation of an oracle. The circuit for this algorithm is the same as the
Deutsch algorithm,
apart from the number of input qubits in the first
register, and is shown in Figure~\ref{fig:deutschstate}.

Assume then that we have a  function $f:\ZZ_2^m\ra \{0,1\}$ which is either balanced or constant.
Again, we employ the standard oracle $U_f$ for $f$.
We use a composite system with an $m$--qubit and a $1$--qubit register.
This time we begin with the basis state $\quzero^{\tens m}\quone$ to which we apply
$W_m\tens W$ to obtain the state
\[\frac{1}{\sqrt{2^m}}\sum_{x=0}^{2^m-1}\ket{x}\tens W\ket{1}. \]
Passing this through the $U_f$ gate  the state of the system becomes
\[\frac{1}{\sqrt{2^m}}\sum_{x=0}^{2^m-1}(-1)^{f(x)}\ket{x}\tens W\ket{1}. \]
Ignoring the second qubit, which is the same for any function, what we have done here is to
encode the function as the coefficients of a {em single} state.
We shall define
$\cD(f)$ to be the
state
\begin{equation}\label{e:deutschstate}
\cD(f)=\frac{1}{\sqrt{2^m}}\sum_{x=0}^{2^m-1}(-1)^{f(x)}\ket{x}.
\end{equation}
 If $f$ is constant then applying
$W_m$ to $\cD(f)$ will give $\pm \quzero$, and if we measure the state we
will always observe $0$.

If $f$ is balanced then we need to make a more careful analysis of the resulting state. In any case,
after applying $W_m$ to $\cD(f)$  we have, using \eqref{e:W_H_n},
\[ \frac{1}{2^m}\sum_{x=0}^{2^m-1}\sum_{y=0}^{2^m-1}(-1)^{f(x)\oplus(x\cdot y)}\ket{y}.
\]
Since $f$ is balanced
the coefficient of $\ket{0}$ in this state is \[\frac{1}{2^m}\sum_{x=0}^{2^m-1}(-1)^{f(x)}=0.\]
Hence, when we measure the state  we {\em never} observe $0$.
 Thus we can distinguish between
balanced and constant functions.

Note that we can only create $\cD(f)$ for
functions $f:\set{0,\ldots,2^{m-1}}\ra\set{0,\ldots,2^{k-1}}$ in the case where $k=1$, but the general idea
is that it is easier to manipulate a single register to gain information about a function than a register
pair containing the standard $2$--register functional superposition provided by the $U_f$ gate. This will also
apply
later, in Shor's algorithm where a measurement of the second register puts the first register into a certain
state which we work with, and also in Grover's algorithm, which manipulates the amplitudes of
$\cD(f)$ 
to ensure that we have a high probability of observing an $x$ for which $f(x)=1$.

In conclusion, we have exhibited a problem which takes time $O(2^m)$ classically
but takes time $O(1)$ on a quantum computer.

The black box query complexity
speed--up is hence exponential, from the classical to quantum setting.
However the exponential speed up is not really robust when compared to
probabilistic
%
algorithms. If one is willing to accept output which is not correct
with absolute
certainty then, given any $\ep>0$, an answer correct with probability $1-\ep$ is
classically
attainable using at most $\log_2(1/\ep)$ queries. Thus, effectively, we have  only
a constant factor speed up, for any fixed permissible probability of error.

In this section we have studied properties of functions which can be deduced exactly
from
$\cD(f)$
with a single quantum measurement.
In a related work \cite{josza}, properties of functions which can be deduced from the standard state
$\sum_{x}\ket{x}\tens\ket{f(x)}$ with a single measurement are studied . Even when a
certain probability of failure is allowed
this class of functions turns out to be very restricted.
\newpage
\section{Shor's algorithm and factoring integers}\label{s:shor}
\subsection{Overview}\label{subs:overview}
Shor's Algorithm, first published in 1994, factors an integer $N$  in  time
$O(L^2\log(L)\log\log(L))$, where $L=\log(N)$.
No classical algorithm of this time complexity
is known.
The problem of factoring integers turns out to be reducible to
that of finding the order of elements of a finite cyclic group. That is,
given an order finding algorithm, an algorithm may be constructed which will, with
high probability return a non--trivial factor of a given composite integer. Moreover
this algorithm may be run quickly on a classical computer. We describe the reduction
of factoring to order finding below.

Order finding in a cyclic group is a special case of the more general problem
of finding the
period of a function which is known to be periodic.
\begin{defn}\label{d:periodic}
If $f$ is a function from a cyclic group $A$ (written additively) to a set $S$ then we say
that $f$ is {\em periodic}, with {\em period} $r\in A$, if the following two conditions hold.
For all $x\in A$
\be[(i)]
\item $f(x)=f(x+r)$ and
\item if $f(x)=f(x+y)$ then $r|y$.
\ee
\end{defn}
Given a periodic function from $\ZZ$ to $\ZZ_N$ the  best known classical algorithm to determine $r$
requires $O(N)$ steps. By contrast, using Shor's algorithm,
\cite{shor1} \cite{shor2}, on a quantum computer only  $O((\log N)^2)$ steps
are required.

As shown by Kitaev \cite{kitaev95}, the essential element of Shor's algorithm is the use of the Quantum Fourier transform
to find an eigenvalue of  a unitary transformation.
This technique was first used
in this way by Simon \cite{simon94} \cite{simon97} to generalise
Deutsch's algorithm. Deutsch's algorithm, Simon's generalisation and
the problems of period finding  can all be regarded  as cases of a problem known
as the
hidden subgroup problem. We shall  describe this problem and various other
applications of the Fourier transform in \ref{subs:hsp}.
Here we first outline the reduction of factoring of integers to order finding, in
Section
\ref{subs:factor_period}. We then describe the essential ingredient to Shor's algorithm,
namely the Quantum Fourier transform and in Section \ref{subs:QFT}. The algorithm itself
is described in Section \ref{subs:rnotintoN} and we end the section with a brief description
of the implementation of the Quantum Fourier transform and an outline of the continued
fractions algorithm which is necessary to extract information from the quantum
part of the main algorithm.
\subsection{Factoring and period finding}\label{subs:factor_period}
%
It is well known that
the ability to find the period of functions effectively would
lead to an efficient algorithm for factoring integers.
In order to see this suppose we wish to factor the integer $N$. We may clearly assume that
$N$ is odd and, since there exist effective probabilistic tests for prime
powers \cite{koblitz},
that $N$ is
divisible by more than one odd prime.
Consider the function $F_N:\ZZ_N\ra\ZZ_N$ given by
\[ F_N(a)=y^a\mbox{ mod }N, \]
where $y$ is a randomly chosen integer in the range $0\lte y<N$. Using the Euclidean
algorithm on $y$ and $N$ we either find a factor of $N$ or we find that $y$ is
coprime to $N$. We may therefore assume that $\gcd(y,N)=1$. If $y$ and $N$ may
both be represented as strings of at most $L$ bits then the total resource required by this step is $O(L^3)$
since this is a bound on the cost of running
the Euclidean algorithm \cite[p. 13]{koblitz}.
With $\gcd(y,N)=1$ the function $F_N$ takes distinct values $1,y,\ldots ,y^{r-1}$, where $r$ is the
(unknown to us) multiplicative order of $y$ modulo $N$. Thus $F$ is periodic of period $r$.
Suppose now that we have computed the multiplicative order $r$ of $y$ modulo $N$, using our hypothetical
period finding algorithm on $F_N$.
Since $N|y^r-1$ the Euclidean algorithm applied to $N$ and $y^r-1$ merely returns the factor $N$ of $N$.
On the other hand, if $r$ is even then
\[1=y^r=(y^{r/2}-1)(y^{r/2}+1) \mod N.\]
As $r$ is minimal with the property that $y^r=1 \mod N$ it follows that $N\nmid y^{r/2}-1$, from which
we see that $N$ and  $y^{r/2}+1$ have a common factor greater than $1$.
We now run the Euclidean algorithm with input $N$ and
$y^{r/2}+1$. If  $N\nmid y^{r/2}+1$ then we obtain a non--trivial factor of $N$, again in time at most
$O(L^3)$.
This step succeeds if and only if we happen to choose $y$ such that $y$ has even order, $r$, and
in addition $N\nmid y^{r/2}+1$. What is the probability of success? Following \cite{ekertjosza} we count
the number of integers $y$ which result in failure. We have $1\le y\le N$ and $\gcd(y,N)=1$, so
that $y\in \ZZ^*_N$ the group of units of $\ZZ_N$, and $r$ is the order of $y$ in this group.
Note
that the order of $\ZZ^*_N$ is $\phi(N)$, where $\phi$ is Euler's totient function:
that is, $\phi(N)$ is the number of integers between $1$ and $N$ which are coprime to $N$ (see \cite[p. 19]{koblitz}
for example).
The notation of  the following lemma is ambiguous, as $|x|$
denotes the order of a group element $x$  and $|X|$ denotes the
cardinality of a set $X$, but the meaning should be clear from the context.
\begin{lemma}[\cite{ekertjosza}]\label{pfacok}
Let $N=p_1^{\a_1}\cdots p_m^{\a_m}$ be the collected prime factorisation of an odd composite integer
$N$ and
let
\[S=\{s\in \ZZ^*_N:|s|\,\textrm{ is odd or }\, s^{|s|/2}=-1\}.\]
Then $|S|\le \phi(N)/2^m$.
\end{lemma}
\begin{pf}
For $n\in \ZZ$ define $l(n)$ to be the greatest integer $d$ such that $2^d|n$.

Consider first the group $\ZZ^*_{p^\a}$, where $p$ is an odd prime and $\a$ a
positive integer. This group is
cyclic of even order $\phi=\phi(p^\a)=p^{\a-1}(p-1)$ with generator $x$, say.
Every element of $\ZZ^*_{p^\a}$ is of the form $x^k$, for some $k\ge 0$, and
for exactly half the elements $k$ is even.
Let $g=x^k$ be an element of $\ZZ^*_{p^\a}$ and  suppose
the order of $g$ is $r$.  If $k$ is even then
\[g^{\phi/2}=x^{k\phi/2}=(x^\phi)^{k/2}=1,\]
so $r|\phi/2$ and it follows that $l(r)<l(\phi)$.
%
Conversely if $k$ is odd then $1=g^r=x^{kr}$ so $\phi |rk$ and,
since $\phi$ is even, this implies $l(\phi)\le l(r)$.
Therefore precisely half of the elements $g$ of $\ZZ^*_{p^\a}$ satisfy
$l(|g|)<l(\phi)$.

Now consider $y\in \ZZ_N^*$.
The Chinese remainder theorem \cite{hardywright} states that
$\ZZ_N^*$ is isomorphic to $\bigoplus^m_{i=1}\ZZ^*_{p_i^{\a_i}}$ under the map taking $y$ to $(y_1,\ldots ,y_m)$, where
$y=y_i\in \ZZ^*_{p_i^{\a_i}}$, for $i=1,\ldots ,m$. Let $y$ have order $r$ in $\ZZ^*_N$ and let $y_i$ have order $r_i$ in
$\ZZ^*_{p^{\a_i}}$. Applying the above isomorphism
we see that $r_i|r$ so $l(r_i)\le l(r)$. First suppose that $r$ is odd. Then $l(r)=0$ so
$l(r_i)=0$, for $i=1,\ldots, m$. Now suppose that $r$ is even but that $y^{r/2} =-1$ in $\ZZ^*_N$. Using
the Chinese remainder theorem again it follows that $y_i^{r/2}=-1$ in $\ZZ_{p^{\a_i}}^*$ and so $r_i\nmid r/2$.
Hence $l(r_i)=l(r)$, for $i=1,\ldots m$. We have shown that if $y\in S$ then $l(r_i)=l(r)$, for
$i=1,\ldots ,m$. To complete the proof we need only show
that this is possible for at most $\phi(N)/2^{m-1}$ elements
of $\ZZ^*_N$.

Fix $y_1\in \ZZ^*_{p_1^{\a_1}}$ and let $r_1=|y_1|$. From the first paragraph of the proof it
follows that there are at most \[\prod_{i=2}^m\frac{\phi(p_i^{\a_i})}{2}\]
elements $(y_2,\ldots,y_m)$ of $\bigoplus_{i=2}^m \ZZ^*_{p_i^{\a_i}}$ such that
$l(r_1)=l(|y_i|)$, for $i=2,\ldots = m$. Summing over all elements of  $\ZZ^*_{p_1^{\a_1}}$,
there are at most
\[\phi(p_1^{\a_1})\prod_{i=2}^m\frac{\phi(p_i^{\a_i})}{2}=\phi(N)/2^{m-1}\]
elements $(y_1,\ldots,y_m)$ of $\bigoplus_{i=1}^m \ZZ^*_{p_i^{\a_i}}$ with $l(|y_1|)=\cdots =l(|y_m|)$.
\end{pf}

Now let $y$ be  an integer chosen uniformly at random
from \[\{y\in \ZZ: 1\le y \le N-1\,\textrm{ and }\,\gcd(y,N)=1\}.\] Then from Lemma \ref{pfacok}
the probability that $y$
has even order in $\ZZ^*_N$ and
$y^{|y|/2}\neq -1 \mod N$
is at least $1-1/2^{m-1}$. The resource cost of the procedure is $O(L^3)$, in addition to the cost of the
order finding routine, and repeating it
sufficiently often we can find a
factor of $N$, with probability as close to $1$ as we like.
Shor's quantum algorithm, described below, finds the order of $y\in \ZZ^*_N$, with probability $1-\ep$, for given
$\ep>0$,  in time polynomial in
$L$. Hence combining Shor's algorithm with the above we have a probabilistic algorithm for factoring integers
which runs in time polynomial in $L=\log(N)$.

\subsection{The quantum Fourier transform}\label{subs:QFT}
The Quantum Fourier Transform is more commonly known (to mathematicians) as the Discrete Fourier Transform.
That is a Fourier transform on a discrete group: which is defined
using characters of irreducible representations.
Since we'll be mainly concerned with Abelian groups we need to know very little about characters, and
we summarise what is necessary here
(for more detail see \cite{jamesliebeck} or \cite{fultonharris}).

A {\em character} of a finite Abelian group $G$ over a field $k$ is a homomorphism from
$G$ to the multiplicative group $k^*$ of non--zero elements of $k$. (More generally a character is the
trace of a representation.)  We shall only
consider characters over $\CC$ here. The set of characters of $G$ is denoted $\hat G$.
Since $G$ is finite, characters must be roots of
unity. In particular the characters of the finite cyclic group $\ZZ_m$ are the
homomorphisms $\chi^c_m=\chi^c$ defined by
\[\chi^c(a) =e^{2\pi i ac/m}, \,\textrm{ where }\, a\in \ZZ_m,\]
for $c=0, \ldots ,m-1$. It's easy to verify that the map $\chi:\ZZ_m\ra\hat\ZZ_m$ defined by
$\chi(a)=\chi^a$ is  an isomorphism from $\ZZ_m$ to $\hat\ZZ_m$. (This result extends to
all finite Abelian groups.)

We shall use the following simple property of characters of cyclic groups.
\begin{lemma}\label{orthog}
  The characters of $\ZZ_m$ satisfy
\be[(i)]
\item ~
  \begin{equation}\label{orthog_form}
    \sum_{a=0}^{m-1}\chi^c(a)=
    \left\{
      \begin{array}{ll}
        m & \,\textrm{ if }\, c=0 \mod m\\
        0 & \,\textrm{ if }\, c\neq0 \mod m
      \end{array}
    \right.
    .
  \end{equation}
\item (Orthogonality of characters)
\begin{equation}\label{orthog_main}
\sum_{a=0}^{m-1}\chi^c(a)\overline{\chi^d(a)}=
    \left\{
      \begin{array}{ll}
        m & \,\textrm{ if }\, c=d \mod m\\
        0 & \,\textrm{ if }\, c\neq d \mod m
      \end{array}
    \right.
    .
\end{equation}
\ee
\end{lemma}
\begin{pf} ~\\[-2em]
\be[(i)]
\item
This is easy to see if $c=0$. If $c\neq 0$ then $\chi^c(1) =e^{2\pi i c/m}$ so $\chi^c(1)\neq 1$ and
is a root of the polynomial $z^m-1=(z-1)(z^{m-1}+\cdots +1)$. The result follows, since
$[\chi^c(1)]^k=\chi^c(k)$.
\item This follows easily from \eqref{orthog_form}, given the observations that
$\overline{\chi^c(x)}=\chi^c(-x)$, $\chi^c(-x)=\chi^{-c}(x)$ and $\chi^c(x)\chi^d(x)=\chi^{c+d}(x)$.
\ee
\end{pf}

For a finite group $G$  the $|G|$--dimensional complex vector space \[\CC G = \bigoplus_{g\in G} \CC \ket{g}\]
is a ring called the {\em complex group algebra},
with multiplication defined  by
\[\left(\sum_{g\in G}a(g)\ket{g}\right)\left(\sum_{h\in G}b(h)\ket{h}\right)=\sum_{g\in G} c(g)\ket{g},\]
where $a(g), b(g)\in \CC $, for all $g\in G$, and $c(g)=\sum_{x\in G} a(x)b(x^{-1}g)$. The group algebra
may also be regarded as the ring of maps from $G$ to $\CC$, the map $a$ sending $g$ to $a(g)$ corresponding
to $\sum_{g\in G} a(g)\ket{g}$. An element of $\CC\ZZ_n$ will be said to be {\em periodic} if it is
periodic as a map from $\ZZ_n$ to $\CC$.

The {\em quantum Fourier transform} on $\ZZ_n$ is a $\CC$--linear map $\QFT=\QFT_n$  from
$\CC \ZZ_n$ to itself defined on the basis vector $\ket{x}$ by
\[\QFT\ket{x}=\frac{1}{\sqrt{n}}\sum_{c=0}^{n-1} \overline{\chi^c(x)}\ket{c},\]
where $\bar{z}$ denotes the complex conjugate of $z$.
Extending by linearity we obtain the image of a general element:
\[ \QFT\left(\sum_{x=0}^{n-1}\alpha(x)\ket{x}\right)=\sum_{c=0}^{n-1}\hat{\alpha}(c)\ket{c}, \]
where
\begin{equation}\label{e:hat}
\hat{\alpha}(c)=\frac{1}{\sqrt{n}}\sum_{x=0}^{n-1}\alpha(x)e^{-\frac{2\pi cxi}{n}}.
\end{equation}

We have seen one instance of the Fourier transform already: the single bit Walsh--Hadamard transformation
$W$ is equal to $\QFT_2$. By comparing coefficients of basis vectors it is easy to see that
$W_n\neq \QFT_n$, for $n\ge 2$, although it is true that $W_n\quzero=\QFT_n\quzero$, for all $n$.

To see that the quantum Fourier transform is a unitary transformation consider its matrix $Q$, relative
to the basis $\ket{x}$, $x\in \ZZ_n$, for $\CC\ZZ_n$. We have $Q=(a_{c,x})$, where the row $c$, column $x$ entry
is
\[a_{c,x}=\frac{1}{\sqrt{n}}
\overline{\chi^c(x)}
, \, 0\le c,x\le n-1.\]
Therefore the conjugate transpose of  $Q$ is $Q^{\dagger}=\left(b_{c,x}\right)$, where
\[b_{c,x}=\frac{1}{\sqrt{n}}
{\chi^x(c)}
,\]
and a straightforward computation using \eqref{orthog_main} shows that $QQ^\dagger=I$.
This shows that $\QFT$ is a unitary transformation and allows us to write down the image
of its inverse, the {\em inverse Fourier transform} $\QFT^{-1}$ on a basis vector $\ket{x}$. That is
\[\QFT^{-1}\ket{x}=\frac{1}{\sqrt{n}}\sum_{c=0}^{n-1} {\chi^c(x)}\ket{c}.\]
In Section \ref{subs:imp_QFT} we shall show how the Fourier transform may be implemented using standard quantum gates.

A crucial property of the quantum Fourier transform is that it
identifies periods. That is,
if $f$ is periodic, with period $r$, and $r$ divides $n$ then
$\QFT(f)$ has non--zero coefficients only at basis vectors $\ket{c}$ which are multiples of $n/r$.
\begin{lemma}\label{aggregate}
Let $f\in \CC\ZZ_n$ and suppose that $f$ is periodic of period $r$, where $r|n$.
Then
\begin{equation}\label{e:aggregate_form}
   \hat{f}(c)=
   \left\{
     \begin{array}{ll}
       \displaystyle
       {
         \frac{\sqrt{n}}{r}\sum_{s=0}^{r-1} f(s)\overline{\chi_n^c(s)},
       }
       & \,\textrm{ if }\, c=0 \mod n/r\\[1.5em]
       0, & \,\textrm{ otherwise }\,
     \end{array}
   \right.
   .
 \end{equation}
\end{lemma}
\begin{pf}
\begin{align*}
\hat{f}(c) & =\frac{1}{\sqrt{n}}\sum_{k=0}^{n-1}f(k)\overline{\chi^c(k)}\\[1em]
& =\frac{1}{\sqrt{n}}\sum_{a=0}^{r-1} \sum_{s=0}^{\frac{n}{r}-1} f(a+sr)\overline{\chi^c(a+sr)}\\[1em]
&=\frac{1}{\sqrt{n}}\sum_{a=0}^{r-1}f(a)\overline{\chi^c(a)}\sum_{s=0}^{\frac{n}{r}-1} \overline{\chi^c(sr)}.
\end{align*}
The result follows on applying Lemma \ref{orthog} modulo $n/r$.
\end{pf}
\subsection{Period finding for beginners}\label{subs:create_periodic}
It is instructive to look first at a restricted case of the period finding algorithm.
Assume that we have a function $f:\ZZ \ra \ZZ_N$ which is periodic of period $r$, as in
Definition \ref{d:periodic}.
In the restricted case we
shall assume that
$N$--ary quantum bits are available, so a basic quantum system is an $N$-dimensional
vector space. Consequently we shall assume, for this section only, that transformations
that we've previously defined using qubits are also defined and implemented
for such $N$--ary quantum bits.
We shall in addition assume that
$r|N$ so that $f$ induces a periodic function from $\ZZ_N$ to itself.
Of course this is an artificial constraint, especially since it is $r$ we're trying to
find, but it illustrates the operation of the algorithm without involving the technical
detail of the general case.

We begin with two registers, of one $N$-ary quantum bit each,
in initial state $\quzero\quzero$. To the first register
 we apply the Walsh--Hadamard transform
to obtain the state \[\frac{1}{\sqrt{N}}\sum_{x=0}^{N-1}\ket{x}\quzero.\]
As in previous sections, we assume that a $U_f$--gate
which evaluates $f$ is available.
We now apply this to our state to
obtain the superposition
\[ \frac{1}{\sqrt N}\sum_{x=0}^{N-1}\ket{x}\ket{f(x)}. \]
At this point we could apply the quantum Fourier transform to the first register of the state.
However it simplifies notation and makes no difference to the outcome if we observe the second register first
(see Example \ref{factor21} below).
So, we now observe
 the second register and obtain, uniformly at random, some value $y_0$ in the image of $f$.
The system then projects to the state
\[ \frac{1}{\sqrt{|f^{-1}(y_0)|}}\sum_{x\in f^{-1}(y_0)}\ket{x}\ket{y_0}. \]
Since $f$ is periodic in the sense of Definition \ref{d:periodic}, there is precisely one value $x_0$ with $0\le x_0<r$ such that
$f(x_0)=y_0$.
If we set $K=N/r$ then, in the first register, we
have  the state
\begin{align*} \ket{\psi} & =\frac{1}{\sqrt K}\sum_{k=0}^{K-1}\ket{x_0+kr}\\
&= \sum_{x=0}^{N-1}\psi(x)\ket{x},
\end{align*}
where
\[\psi(x)=\left\{
  \begin{array}{ll}
    \displaystyle\frac{1}{\sqrt{K}}=\sqrt{\frac{r}{N}}, &\textrm{ if }\, r|x-x_0\\[1em]
    0, &\textrm{ otherwise}
  \end{array}
\right.
.
\]
Thus $\psi\in \CC\ZZ_N$ is periodic with period $r$.
\begin{figure}
\begin{center}
\psfrag{x}{$\ket{x}$}
\psfrag{f}{$\ket{f(x)}$}
\psfrag{r}{$r_2\leadsto\ket{2}$}
\includegraphics[width=3.6in]{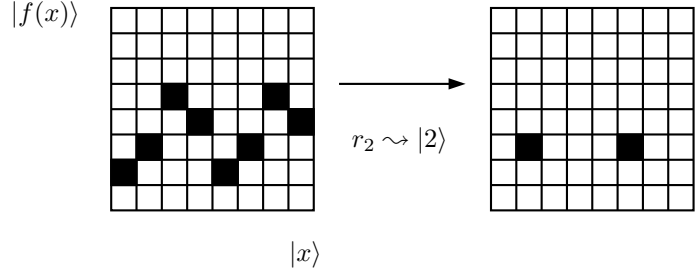}
\caption{\label{fig:periodicstate} Collapsing to a periodic state}
\end{center}
\end{figure}
For example, for the function drawn in Figure \ref{fig:periodicstate}, where $N=8$ and
$r=4$, we have observed $\ket{2}$ in the second register and this
has set the value $x_0$ to $1$.
Notice that, since $y_0$ is a random element of the image of $f$,
observation of the first register at this stage returns a
value $x_0+kr$, for some uniformly random  $k\in\{0,\ldots,N/r-1\}$. That is, observation of
the periodic state $\ket{\psi}$ will simply give a uniformly random value in
$\{0,\ldots,N-1\}$ and yields
no information on the value of $r$ at all!

Now, if we apply $\QFT$ to the first register then, using Lemma \ref{aggregate},
we obtain
\[\QFT\ket{\psi} = \sum_{c=0}^{N-1} \hat{\psi}(c)\ket{c},\]
where
\begin{align*}
\hat{\psi}(c)& =
\left\{
 \begin{array}{ll}
    \displaystyle\frac{\sqrt{N}}{r}\sum_{s=0}^{{r}-1}\psi(s)\overline{\chi^c(s)},& \mbox{if $c= 0\mod\frac{N}{r}$} \\[2em]
    0, & \mbox{otherwise}
  \end{array}
\right.
\\[1em]
& =
\left\{
 \begin{array}{ll}
    \displaystyle\frac{1}{\sqrt{r}}\overline{\chi^c(x_0)},& \mbox{if $c= 0\mod\frac{N}{r}$} \\[2em]
    0, & \mbox{otherwise}
  \end{array}
\right.
,
\end{align*}
since $\psi(x_0)=1/\sqrt{K}$ and $\psi(s)=0$ for all $s\neq x_0$.
Therefore, setting $c(s)=sN/r$,
\begin{equation}\label{e:easy_output}
\QFT\ket{\psi}=\frac{1}{\sqrt{r}}\sum_{s=0}^{r-1}\overline{\chi^{c(s)}(x_0)}\ket{c(s)} .
\end{equation}
Now we observe this state. We obtain a value $c=c(s)$ which is a multiple of $N/r$. In fact,
for a uniformly
random $s\in\{0,\ldots,r-1\}$ we have
\[ c=s \frac{N}{r}\implies \frac{c}{N}=\frac{s}{r} \]
where the fraction $c/N$ is known. If we reduce $c/N$ down to lowest terms (using the Euclidean
algorithm) then we may determine $r$, as the denominator of this irreducible fraction, provided that
$\gcd(s,r)=1$. If $s$ and $r$ are not coprime then we will obtain a proper factor
of $r$ and not $r$ itself. To see that this is not really a problem, we appeal to a result from number theory
(see \cite{hardywright} for example):
\[ \liminf\left(\frac{\phi(n)}{n/\log_e\log_e n}\right)=e^{-\gamma}
\]
where $\gamma$ is a constant known as {\em Euler's constant}.
This means that if we choose a random number from $\{0,\ldots,n-1\}$ then the probability $p(n)$ that it is coprime to
$n$ satisfies
\begin{equation}\label{e:cpn}
p(n)=\frac{\phi(n)}{n}\gte \frac{e^{-\gamma}}{\log_e\log_e n} .
\end{equation}
So  we obtain a number coprime to $N$ with probability
$1-\ep$, where $\ep>0$ can be made arbitrarily small, by repeating the
above observation $O(\log_e\log_e N)$ times.

In summary, at each iteration the process outputs a number $c$.
A value $r$ may then be read off from the equality $c/N=s/r$, where $\gcd(r,s)=1$.
Repeating the process sufficiently many times we may compute the period of $f$,
to within a given probability, as
the least common multiple of the non--zero $r$ values.

\begin{ex}\label{factor21}
Suppose we wish to factorise $21$. We begin by choosing an integer coprime to $21$, say $4$, as described in Section
\ref{subs:factor_period}. The function $f:\ZZ_{21}\ra \ZZ_{21}$ is given by $f(k)=4^k$ and has period $r$, which we
wish to find. After applying $W_{21}$ and $U_f$ we have the state
\begin{align*}
\frac{1}{\sqrt{21}}
[
&(\ket{0}+\ket{3}+\ket{6}+\ket{9}+\ket{12}+\ket{15}+\ket{18})\ket{1}\\
+&(\ket{1}+\ket{4}+\ket{7}+\ket{10}+\ket{13}+\ket{16}+\ket{19})\ket{4}\\
+&(\ket{2}+\ket{5}+\ket{8}+\ket{11}+\ket{14}+\ket{17}+\ket{20})\ket{16}
].
\end{align*}
Observing the second register we obtain, with probability $1/3$, one of $1$, $4$ or $16$. The first register
then contains $\ket{\psi_0}$,  $\ket{\psi_1}$ or $\ket{\psi_2}$, where
$\ket{\psi_s}=\sum_{k=0}^{6}\ket{s+3k}$.

Applying $\QFT_{21}$ to these states we have
\begin{align*}
\QFT\ket{\psi_0} & = \frac{1}{\sqrt{3}}(\quzero+\ket{7}+\ket{14}) \,\textrm{ or }\\
\QFT\ket{\psi_1} & = \frac{1}{\sqrt{3}}(\quzero+\w\ket{7}+\w^2\ket{14}) \,\textrm{ or }\\
\QFT\ket{\psi_2} & = \frac{1}{\sqrt{3}}(\quzero+\w^2\ket{7}+\w\ket{14}),\\
\end{align*}
where $\w=e^{-2\pi i/3}$. Whichever of these we have, observation now yields, with equal probabilities,
$c=0$, $c=7$ or $c=14$. If we observe $c=0$ then the process must be run again. If we observe $c=7$ then
$c/N=7/21=1/3$ and the denominator of this fraction is $r$. Similarly, if $c=14$ we read $r$ off from
$c/N=14/21=2/3$. (Of course the example has been set up in the knowledge that the order of $4$ divides
$21$, merely for purposes of illustration. We should not be misled to the conclusion that the example
generalises to a simple method for factoring integers.)

Notice that if we had omitted the observation of the second register before application of $\QFT$ then, applying
$\QFT$ would have resulted in the state
\[\frac{1}{3}[(\quzero+\ket{7}+\ket{14})\ket{1}+(\quzero+\w\ket{7}+\w^2\ket{14})\ket{4}
+(\quzero+\w^2\ket{7}+\w\ket{14})\ket{16}]
\]
and observation would have given the same result as before.
\end{ex}

\subsection{Advanced period finding}\label{subs:rnotintoN}
There are two immediate problems to be overcome in implementation of the algorithm in general.
First of all
we have defined quantum computation in terms of qubits, not $N$--ary bits. In this setting we
need at least $L=\lceil \log(N)\rceil$ qubits to represent $N$ as an binary integer.
This means that we shall
have to run the algorithm using the quantum Fourier transform $\QFT_q$
for some $q$ which is not equal to
$N$.
Secondly,
we cannot assume that the period $r$ divides $q$ (or $N$).
If $q/r$ is not an integer then a periodic function $f:\ZZ\ra \ZZ_N$ of period $r$
does not induce a well--defined
function from $\ZZ_q$ to $\ZZ_N$.

Suppose then that we have a periodic function $f:\ZZ\ra \ZZ_N$ of period $r$, as before, but
now we do not assume $r|N$. To address
the first problem above we choose $q=2^n>N$, for some integer $n$. We shall see below that the
success of the algorithm depends on making a good choice for $q$.
Recall that we identify $a_0\tens\cdots\tens a_{n-1}\in\ZZ_2^{\tens n}$ with a binary integer
via the map $b$ such that $b(a_0\tens\cdots\tens a_{n-1})=2^{n-1}a_0+\cdots a_{n-1}$.
We use the composite $f\circ b:\ZZ_2^{\tens n}\ra \ZZ_N$ to simulate $f$.
Replacing $f$ by this function observe that it satisfies the conditions
that, for all $x$ such that $0\le x<q-r$,
\be[(i)]
\item $f(x)=f(x+r)$ and
\item $f(x+y)=f(x)$ implies $r|y$, if $0\le y<q-r-x$.
\ee
We shall also use the map $b$ to identify $\ZZ_2^{\tens n}$ with the
set $\{0,\ldots ,q-1\}$, which we regard as $\ZZ_q$. In particular this identification
allows us to apply the Fourier transform $\QFT_q$ to $\ZZ_2^{\tens n}$.

\begin{figure}
\begin{center}
\psfrag{exact}{exact case}
\psfrag{non}{non--exact case}
\psfrag{0}{$0$}
\psfrag{N}{$\frac{q}{r}$}
\psfrag{2N}{$\frac{2q}{r}$}
\psfrag{3N}{$\frac{3q}{r}$}
\psfrag{QFT}{$\QFT$}
\psfrag{x0}{$x_0$}
\psfrag{x0r}{$x_0+r$}
\psfrag{x02}{$x_0+2r$}
\includegraphics[width=3.8in]{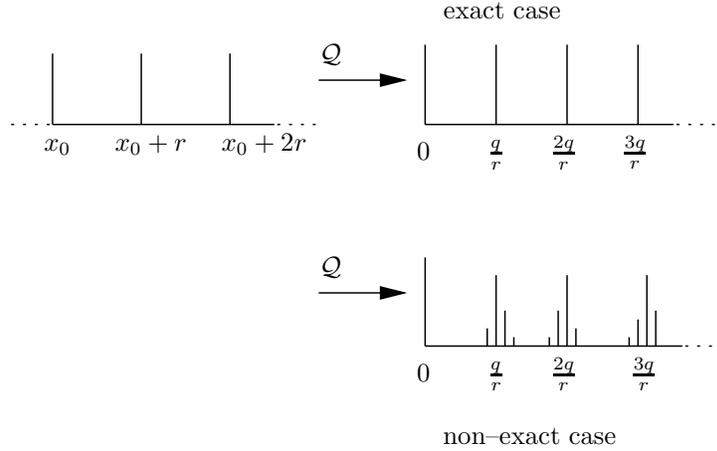}
\caption{\label{fig:qft} Exact and non--exact cases for the quantum Fourier transform}
\end{center}
\end{figure}

We can set up our algorithm, along the lines of Section \ref{subs:create_periodic}, starting with two registers,
the first of $n$ qubits and the second of $L$ qubits in initial state $\quzero\quzero$.
As before we apply Walsh--Hadamard $W_n$ and
a $U_f$ gate to obtain the state
\begin{equation} \label{e:pstate}
\frac{1}{\sqrt q}\sum_{x=0}^{q-1}\ket{x}\ket{f(x)}.
\end{equation}
However, if
$q/r$ is not an integer then after applying
the quantum Fourier transform $\QFT_q$ to the first register of
\eqref{e:pstate} there may be non--zero coefficients
$\hat{\psi}(c)$ at values of $c$ which are not multiples of $q/r$.
Nevertheless it turns out that
the non--zero coefficients are clustered around points close to multiples of $q/r$
(see Figure~\ref{fig:qft}).
This means that, if a good choice of $q$ is made, there is a high probability of
an observation returning a value close enough to one of these multiples
to allow the
use of a classical algorithm, based on properties of continued fractions,
to yield the value of $r$.

First we shall state the required property of continued fractions and describe how it can be used to
extract information from the algorithm. Then we'll complete the description of the algorithm.

The property we need of continued fractions is the following. The necessary definitions and background can be
found in Section \ref{subs:cont_frac}.
\begin{theo}{\rm\cite{hardywright}}\label{th:contfrac}
If $x\in\QQ$ satisfies
\[\left| \frac{p}{r}-x \right|<\frac{1}{2r^2}\]
then ${p}/{r}$ is a convergent of the continued fraction expansion of $x$.
\end{theo}
(This theorem holds for all $x\in\RR$, but we do not need this here and we've only defined continued fractions
for rational numbers.)

To see how this helps suppose that, as in Section \ref{subs:create_periodic}, we apply $\QFT_q$ to the first register
of \eqref{e:pstate} and then observe to obtain a value $c$. If
\[\left|c-s\frac{q}{r}\right|=q\left|\frac{c}{q}-\frac{s}{r}\right|<\frac{q}{2r^2},\]
for some integer $s$ then Theorem \ref{th:contfrac} implies that
$s/r$ is a convergent of the continued fraction expansion of $c/q$. As $c$ and $q$ are known these
convergents may all be calculated using the Euclidean algorithm. If, in addition,
$\gcd(s,r)=1$ then the denominator of one of the convergents is $r$.

Furthermore, given $s$ with $0\le s<r$ there is a unique integer $c_s$ such
that $-r/2\le c_sr-sq<r/2$ and
for such $c_s$ we have
\begin{equation}\label{e:bound2q}
\left|\frac{c_s}{q}-\frac{s}{r}\right|<\frac{1}{2q}.
\end{equation}
This motivates the choice of $q$ as the unique integer $q=2^n$ such that
$N^2\le q< 2N^2$: for then, with $s$ and $c_s$ as above,
\begin{equation}\label{e:boundedc}
\left|\frac{c_s}{q}-\frac{s}{r}\right|\le\frac{1}{2q}<\frac{1}{2N^2}<\frac{1}{2r^2},
\end{equation}
since $r<N$.

Given this choice of $q$,
if we observe $c_s$ for $s$ such that $\gcd(r,s)=1$, then
we can compute the convergents of the continued fraction expansion
of $c_s/q$ using the continued fraction algorithm, as described in
Section \ref{subs:cont_frac}. From Theorem \ref{th:contfrac} and \eqref{e:boundedc},
$s/r$ is among these convergents.
It is a further consequence of our choice of $q$ that $s/r$
is the unique convergent of $c_s/q$ satisfying the inequality \eqref{e:bound2q}. To see this
suppose that $a$ and $b$ are positive integers with $b<N$ such that $a/b$ also satisfies
\eqref{e:bound2q}. Then
\[
\left|\frac{s}{r}-\frac{a}{b}\right| \le \left|\frac{s}{r}-\frac{c}{q}\right|+
\left|\frac{c}{q}-\frac{a}{b}\right|<\frac{1}{q}.
\]
This implies that $|sb-ar|<rb/q$ and as $N^2\ge q$ it follows that $|sb-ar|<1$ so that
$s/r=a/b$.
Therefore, as claimed, $s/r$ is the unique convergent satisfying \eqref{e:bound2q}.
This being the case we can use \eqref{e:bound2q} to find $s/r$
amongst the convergents of $c_s/q$ and this allows us to compute
$r$. The time taken to do this, once $c_s$ has been observed, is therfore $O(L^3)$.
Therefore we shall need to know the probability of observing $c_s$, for $s$ such that $\gcd(r,s)=1$.

To start with we observe the second register of \eqref{e:pstate} and obtain some value
$b\in \ZZ_N$. As before, there is  $a$ such that $f(a)=b$ and $0\le a<r$, and so
$f^{-1}(b) =\{a+kr: 0\le k<K_a\}$, where $K_a$ is the greatest integer such that $(K_a-1)r+a<q$.
In the first register we now have
  \begin{align*} \ket{\psi} & =\frac{1}{\sqrt K_a}\sum_{k=0}^{K_a-1}\ket{a+kr}\\
 &=\sum_{x=0}^{q-1}\psi(x)\ket{x},
\end{align*}
where
\[\psi(x)=\left\{
  \begin{array}{ll}
    \displaystyle\frac{1}{\sqrt{K_a}}, &\textrm{ if }\, r|x-a\\[1em]
    0, &\textrm{ otherwise}
  \end{array}
\right.
.
\]
We now apply $\QFT$ to the first register and obtain
\begin{equation}\label{e:qperiod}
\QFT\ket{\psi}=\sum_{c=0}^{q-1} \hat{\psi}(c)\ket{c},
\end{equation}
where, using  \eqref{e:hat} and an arguement similar to that of the proof of Lemma \ref{aggregate},
\begin{equation}\label{e:phat}
\hat{\psi}(c)=\frac{1}{\sqrt{K_a q}}\sum_{k=0}^{K_a-1}e^{-2\pi ic(a+kr)/q}.
\end{equation}
We now observe the first register and use the following estimate, which we prove
in Section \ref{subs:Probability_estimates}, following
\cite{ekertjosza} and \cite{hirvensalo}.

\begin{prop}\label{prop:esitmate}
 The probability of observing \eqref{e:qperiod} and obtaining a value $c_s$
such that \[\left| \frac{c_s}{q}-\frac{s}{r}\right|<\frac{1}{2r^2}\] and $\gcd(s,r)=1$ is at least
\[\frac{4\phi(r)}{\pi^2 r^2}\left(1-\left(\frac{\pi r}{2 q}\right)^2\right).\]
In particular, if $N\ge 158$ and $r\ge 19$  then this probability is at least $1/10\log_e\log_e(N)$.
\end{prop}
\noindent{\bf Summary: the period finding algorithm}~\\[.5em]
Given a function $f:\ZZ\ra \ZZ_N$ which is periodic of period $r$ perform the following steps.
\be[{\bf (1)}]
\item First check whether $f$ has period $r<19$. The number of operations this requires depends on $f$. For example
if $f$ is modular exponentiation, as in Shor's algorithm, then then this step requires $O(log^3(N))$ operataions.
\item Compute $L=\lceil \log(N)\rceil$ and set $q=2^n$, where $2\log(N)\le n<2\log(N)+1$.
This may be done using  a classical algorithm in $O(L)$ operations.
\item\label{it:s2} Prepare first and second registers $Q$ and $R$ of $n$ and $L$ qubits, respectively,
in state $\quzero\quzero\in Q\tens R$.
\item\label{it:s3} Apply $W_n\tens I_L$ to the state of \stref{it:s2}. The Walsh--Hadamard transformation $W_n$
may be implemented using $n$ single qubit Walsh--Hadamard gates so the number of operations required in this
step is $O(n)=O(L)$.
\item In this step we assume the existence of a unitary transformation $U_f$ from
$Q\tens R$ to itself, which maps basis vector $\ket{x}\ket{y}$ to
$\ket{x}\ket{f(x)\oplus y}$. Apply $U_f$ to the output of \stref{it:s3} to give
\eqref{e:pstate}. The complexity of this step is dependent on $f$. For example if $f$ is modular exponention then
$U_f$ may be implemented using $O(L^3)$ operations.
\item Observe the second register of \eqref{e:pstate} and project to a state
$\ket{\psi}\ket{b}$, where $\ket{\psi}\in Q$ and $b\in \ZZ_N$.
\item\label{it:s6} Apply $\QFT_q\tens I_L$ to $\ket{\psi}\ket{b}$ to obtain \eqref{e:qperiod} in the
first register (and $\ket{b}$ in the second). We show in Section \ref{subs:imp_QFT} that $\QFT_q$ may be
implemented in $O(n^2)=O(L^2)$ operations.
\item Observe the state of \stref{it:s6} and obtain a basis vector $\ket{c}\ket{b}$.
\item\label{it:s8} Use the continued fraction algorithm (see Section \ref{subs:cont_frac}) to
find the convergents of $c/q$ and output a candidate $r^\prime$ for the period of $f$.
This requires $O(L^3)$ operations.
%
\ee
Modulo the complexity of $f$ the above procedure requires $O(L^3)$ operations.
>From Proposition \ref{prop:esitmate}
the final step is succesful with probability $1/10\log_e\log_e(N)$.
Hence we repeat the algorithm
$O(\log\log(N))=O(log(L))$ times and compute the least common multiple
$l$ of the non--zero values $r^\prime$ output in the last step.
Then with high probability, $l=r$, the period of $f$. Hence we have computed $r$
in $O(L^3\log(L))$ operations,
subject to the time required to compute $f$ and the probability of error.
In the case of Shor's algorithm, where $f$ is exponentiation modulo $N$, it follows that we can perform quantum
modular exponentiation in time $O(L^3\log(L))$, where $L=\log(N)$. As the reduction of factoring to period finding
described in Section \ref{subs:factor_period} requires $O(L^3)$ operations we have a
quantum algorithm for
factoring an integer $N$ in time $O(L^3\log(L))$. In fact this bound is not tight and, by using
faster algorithms for integer arithmetic,
Shor \cite{shor2} obtains the bound of $O(L^2\log(L)\log\log(L))$
mentioned at the begining of this
section.
%
\subsection{Probability estimates}\label{subs:Probability_estimates}
We shall need the following result from \cite{rosserschoenfeld}.
\begin{theo}\label{rosser}
For $r\ge 3$,
\[\frac{r}{\phi(r)}<e^{\gamma}\log_e\log_e(r)+\frac{2.50637}{\log_e\log_e(r)},\]
where $\gamma=\lim_{n\ra \infty}(1+\frac{1}{2}+\frac{1}{3}+\cdots +\frac{1}{n}-\log_e(n))=
0.57721556649\ldots$ is Euler's constant.
\end{theo}
Recall from Section \ref{subs:rnotintoN} that $f$ is a periodic function from $\ZZ_q$ to $\ZZ_N$ of period $r$, that
$L,n,N,r,q\in \ZZ$, with $1\le r<N$, $q=2^n$,
$N^2\le q<N$ and $L=\log\lceil N \rceil$. We have a $2$ register quantum system with first and second
registers of $n$ and $L$ qubits, respectively. The first register is in the state given by \eqref{e:qperiod} and
the second in state $\ket{b}$, where $b=f(a)$. In addition $K_a$ is the largest integer such
that $(K_a-1)r+a<q$.
In the special case of Section \ref{subs:create_periodic} we
measured \eqref{e:easy_output} and, with high probability, observed $c$ such that $cr-Ns=0$, for some
$s$ with $\gcd(r,s)=1$.
In the general case we shall show that when we measure \eqref{e:qperiod} there is a high probability of
observing $c$ such that $-r/2\le cr-sq<r/2$, for some integer $s$ such that $\gcd(r,s)=1$.
This will be enough to allow us to compute $r$.
\begin{pf}[Proof of Proposition \ref{prop:esitmate}]
Recall that, given $s\in \ZZ$, we write $c_s$ for the unique integer such that $-r/2\le c_s r -sq<r/2$.
Simple calculations show that $0\le c_s <q$ if and only if $0\le s< r$ and also that if $c_s=c_t$ then
$s=t$. Hence the integers $c_0,\ldots ,c_{r-1}$ are distinct and lie in $[0,q)$.

For each $s$ with $0\le s<r$ write $\e_s=c_sr-sq$ and define
\[\q_s=\frac{2\pi\e_s}{q}.\]
The conditions on $K_a$ imply that
\begin{equation}\label{e:K_cond}
\frac{q}{r}-1<K_a<\frac{q}{r}+1.
\end{equation}
Therefore, for $s$ such that $0\le \e_s<r/2$ and for $k$ such that $0\le k<K_a$,  we have
\[0\le k\q_s<\frac{\pi k r}{q}<\pi .\]
Hence, for such $s$, the points $e^{-ik\q_s}$ lie in the lower half--plane, for $k=0,\ldots ,K_a-1$
(see Figure \ref{fig:circle1}).
\begin{figure}[!btph]
\begin{center}
\includegraphics[scale=1]{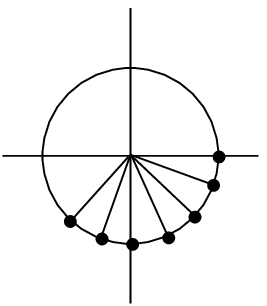}
\caption{\label{fig:circle1} The points $e^{-ik\q_s}$, for $s$ such that $0\le \e_s<r/2$.}
\end{center}
\end{figure}
Similarly, for  $s$ such that $-r/2\le \e_s<0$ the points
$e^{-ik\q_s}$ lie in the upper half--plane, for $k=0,\ldots ,K_a-1$.

It follows (see Figure \ref{fig:circle2}) that for all $s$ such that $0\le s<r$
\begin{equation}\label{e:const_int}
\left|\sum_{k=0}^{K_a-1}e^{-i\q_s k}\right|\ge\left|\sum_{k=0}^{K_a-1} e^{\frac{-i\pi r k}{q}}\right|.
\end{equation}
This is known as {\em constructive interference}.
\begin{figure}[!btph]
\begin{center}
\psfrag{q}{$\q_s$}
\psfrag{p}{$\frac{\pi r }{q}$}
\includegraphics[scale=1]{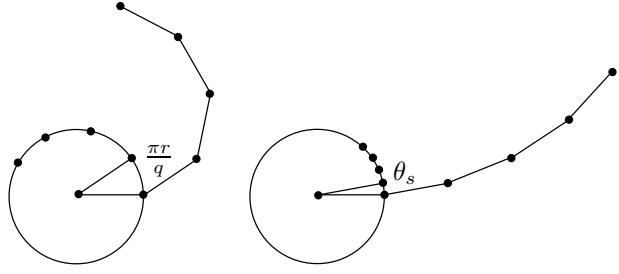}
\caption{\label{fig:circle2} Different rates of constructive interference}
\end{center}
\end{figure}

We shall denote by $p_1(c)$ the probability that $c$ is observed when the first register is measured
with respect to the computational basis.
The state \eqref{e:qperiod} has coefficients given by \eqref{e:phat} so
\begin{align}
\nonumber
p_1(c)&=
\frac{1}{K_a q}\left|\sum_{k=0}^{K_a-1}e^{-\frac{2\pi ic(kr+a)}{q}}\right|^2\\[.5em]
\nonumber
&=
\frac{1}{K_a q}\left|e^{-\frac{2\pi ica}{q}}\sum_{k=0}^{K_a-1}e^{-\frac{2\pi ickr}{q}}\right|^2\\[.5em]
&=
\frac{1}{K_a q}\left|\sum_{k=0}^{K_a-1}e^{-\frac{2\pi ickr}{q}}\right|^2.\label{e:pc}
\end{align}

Since
\[
\frac{-2\pi i c_s r}{q} = \frac{-2\pi i(sq + \e_s)}{q} = -2\pi i s - \frac{2\pi i \e_s}{q},
\]
we have
\begin{equation}\label{e:phi}
e^{(-2\pi i c_s r)/q}=e^{(2\pi i \e_s)/q}=e^{-i\q_s k},
\end{equation}
for $0\le s<r$.

Combining \eqref{e:const_int}, \eqref{e:pc} and \eqref{e:phi} we have
\begin{align}
\nonumber
p_1(c) & \ge \frac{1}{K_a q}\left|\sum_{k=0}^{K_a-1} e^{-(i\pi r k )/q}\right|^2\\[.5em]
\nonumber
& = \frac{1}{K_a q}\left| \frac{\left(e^{-(i\pi r  )/q}\right)^{K_a}-1}{e^{-(i\pi r  )/q}-1}\right|^2\\[.5em]
&= \frac{\sin^2(\pi r K_a/2q)}{K_a q \sin^2(\pi r/2q)},\label{e:psin}
\end{align}
using the identity $\left|e^{ix}-1\right|^2=4\sin^2(x/2)$, for $x\in \RR$.

>From \eqref{e:K_cond} we have
\[
0<\frac{\pi}{2}\left(1-\frac{r}{q}\right)<\frac{K_a\pi r}{2q}<\frac{\pi}{2}\left( 1+\frac{r}{q}\right)<\pi,
\]
so
\begin{equation}\label{e:sin1}
\sin\left( \frac{K_a\pi r}{2q}\right)>\sin\left(\frac{\pi}{2}\left( 1+(-1)^d\left(\frac{r}{q}\right)\right)\right),
\end{equation}
for $d=1$ or for $d=-1$. As $r/q$ is small we have (using  a Taylor series expansion)
\begin{equation}\label{e:sin2}
\sin^2\left( \frac{\pi}{2}\left( 1\pm \frac{r}{q}\right)\right)\ge 1-\left(\frac{\pi r}{2 q}\right)^2
\end{equation}
and
\begin{equation}\label{e:sin3}
\sin(\pi r/2q)\le \pi r/2 q.
\end{equation}

Combining \eqref{e:psin}, \eqref{e:sin1}, \eqref{e:sin2} and \eqref{e:sin3} we see
that the probability of observing $c_s$, for some $s$ such that $0\le s<r$, is
\begin{align}
\nonumber
p_1(c_s) &\ge \frac{1}{K_a q}\left(1-\left( \frac{\pi r}{2q}\right)^2\right)\left( \frac{2q}{\pi r}\right)^2\\[.5em]
\nonumber
&= \frac{q}{K_a}\left(1-\left( \frac{\pi r}{2q}\right)^2\right)\left( \frac{2}{\pi r}\right)^2\\[.5em]
\nonumber
& > \frac{rq}{r+q}\left(1-\left( \frac{\pi r}{2q}\right)^2\right)\left( \frac{2}{\pi r}\right)^2\\[.5em]
&\ge \frac{4}{\pi^2 r}\left(1-\left( \frac{\pi r}{2q}\right)^2\right).\label{e:pgoodc}
\end{align}
As discussed in Section \ref{subs:create_periodic} the probability that $s\in
\{0,\ldots ,r-1\}$ is coprime to $r$ is $\phi(r)/r$. Together with
\eqref{e:boundedc} and
\eqref{e:pgoodc}
this yields the first statement of the proposition.

For the final statement we note that if $N\ge 158$ then
\[
\frac{4}{\pi^2}\left(1-\left( \frac{\pi r}{2q}\right)^2\right)\ge \frac{2}{5}.
\]
Since there are $r$ distinct integers, $c_0, \ldots ,c_{r-1}$ this implies
that $p_1(c_s)\ge 2/5$. From Theorem \ref{rosser} it follows that if $r\ge 19$
then
\[
\frac{\phi(r)}{r}>\frac{1}{4\log_e\log_e(r)}>\frac{1}{4\log_e\log_e(N)},
\]
and the final statement of the proposition follows.
\end{pf}
\subsection{Efficient implementation of the quantum Fourier transform}\label{subs:imp_QFT}

As in the case of the Deutsch--Jozsa algorithm it is the complexity of Shor's algorithm which is
of interest. Shor showed that his algorithm could be implemented
efficiently using small quantum circuits because this is also true of  the quantum Fourier transform
on $\ZZ_{2^n}$.
In fact this implementation of the quantum Fourier transform is essentially an adaptation of the
standard fast Fourier transform technology to quantum computation.
Here we'll show that it is possible to implement the quantum Fourier transform on $\ZZ_n$ using at most $n^2$
$2$--qubit gates.

\begin{figure}[!btph]
\begin{center}
\psfrag{U}{$U$}
\includegraphics[width=1.8in]{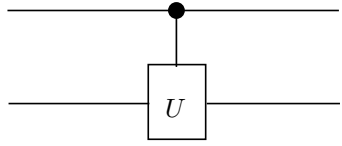}
\caption{\label{fig:controlled} A controlled $U$--gate}
\end{center}
\end{figure}

If $U$ is a single qubit quantum gate, then we define the {\em controlled }$U${\em --gate}, a unitary transformation
of a $2$--qubit system, by
\[ \quzero\langle 0|\tens I + \quone\langle 1|\tens U. \]
If the first qubit is $\quone$ then $U$ is applied to the second qubit. If it is $\quzero$ then the identity $I$ is
applied to the second qubit. The controlled $U$--gate is depicted in a quantum circuit diagram in Figure~\ref{fig:controlled}.

\begin{figure}[!btph]
\begin{center}
\psfrag{W}{$W$}
\psfrag{B1}{$B_1$}
\psfrag{B2}{$B_2$}
\psfrag{B3}{$B_3$}
\psfrag{k3}{$\ket{k_3}$}
\psfrag{k2}{$\ket{k_2}$}
\psfrag{k1}{$\ket{k_1}$}
\psfrag{k0}{$\ket{k_0}$}
\psfrag{b3}{$\ket{b_3}$}
\psfrag{b2}{$\ket{b_2}$}
\psfrag{b1}{$\ket{b_1}$}
\psfrag{b0}{$\ket{b_0}$}
\psfrag{p1}{{\small phase $1$}}
\psfrag{p2}{{\small phase $2$}}
\psfrag{p3}{{\small phase $3$}}
\psfrag{p4}{{\small phase $4$}}
\includegraphics[width=4.5in]{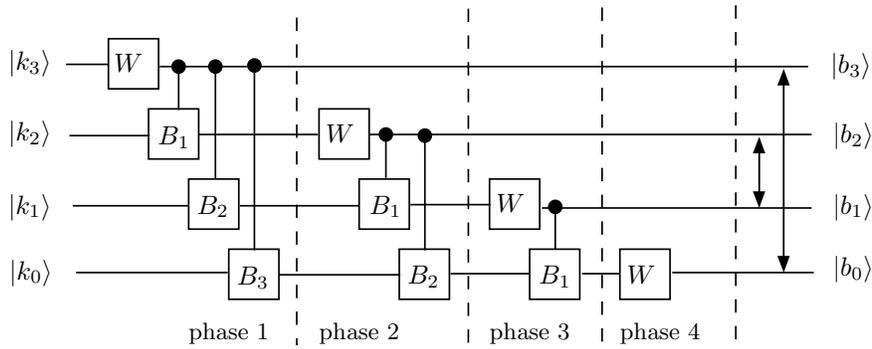}
\caption{\label{fig:4qubitqft} A quantum circuit for the $4$--qubit quantum Fourier transform}
\end{center}
\end{figure}

Let $R_\phi$ be the unitary transformation (called a {\em phase shift})
on a single qubit given by the matrix
\[ \left( \begin{array}{ll} 1 & 0 \\ 0 & e^{-i\phi} \end{array}
\right)
\]
Let $B_k$ ($k\in\NN$) denote $R_{\pi/2^k}$. Now
consider the $4$--qubit circuit in Figure~\ref{fig:4qubitqft}, which is
constructed using controlled $B_k$--gates. We claim that this performs the quantum Fourier transform
on $\ZZ_{2^4}$.
The first Walsh--Hadamard gate applies $W\tens I^{\tens 3}$ to $\ket{k_3k_2k_1k_0}$ and results in
\[
W\tens I^{\tens 3}\ket{k_3k_2k_1k_0}=
\left\{
\begin{array}{ll}
\frac{1}{\sqrt{2}}(\quzero+\quone)\tens |k_2k_1k_0\rangle & \mbox{ if $k_3=0$} \\
\frac{1}{\sqrt{2}}(\quzero-\quone)\tens |k_2k_1k_0\rangle & \mbox{ if $k_3=1$}
\end{array}\right.
.
\]
This can be written as
\[
\frac{1}{\sqrt{2}}(\quzero+e^{-\pi i k_3}\quone)\tens |k_2k_1k_0\rangle.
\]
The map corresponding to the first controlled--$B_1$ gate is
\[
\quzero\langle 0|\tens I^{\tens 3}+\quone\langle 1|\tens R_{\pi/2}\tens I^{\tens 2}.
\]
After passing through this gate the quantum state becomes
\[
\left\{
\begin{array}{ll}
\frac{1}{\sqrt{2}}\left(|00k_1k_0\rangle+e^{-\pi i k_3}|10k_1k_0\rangle\right) & \mbox{ if $k_2=0$} \\
\frac{1}{\sqrt{2}}\left(|01k_1k_0\rangle+e^{-\pi i k_3}\quone\tens e^{-\pi i/2}\quone\tens |k_1k_0\rangle\right) &
\mbox{ if $k_2=1$}
\end{array}
\right.
\]
which can be written as
\[ \frac{1}{\sqrt{2}}\left(\quzero+e^{-\pi i\left(k_3+\frac{k_2}{2}\right)}\quone\right)\tens |k_2k_1k_0\rangle.
\]
Using similar calculations, after phase $1$ the state becomes
\[
\frac{1}{\sqrt{2}}\left(\quzero+e^{-\pi i\left(k_3+\frac{k_2}{2}+\frac{k_1}{4}+\frac{k_0}{8}\right)}\quone\right)
\tens |k_2k_1k_0\rangle
.
\]
That is, we have
\[ |k\rangle\mapsto\frac{1}{\sqrt{2}}\left(\quzero+e^{-\frac{2\pi ki}{16}}\quone\right)
\tens |k_2k_1k_0\rangle,
\]
where $|k\rangle=|k_3k_2k_1k_0\rangle$, and we can further rewrite the right hand side as
\[
\frac{1}{\sqrt{2}}\sum_{b_0=0}^1e^{-\frac{2\pi i k b_0}{16}}|b_0k_2k_1k_0\rangle
.
\]
Similarly, after phase $2$ we obtain
\[
\frac{1}{\sqrt{2^2}}\sum_{b_0=0}^1\sum_{b_1=0}^1e^{-\frac{2\pi i(b_0+2b_1)}{16}}|b_0b_1k_1k_0\rangle
\]
and after phase $4$ the state becomes
\[
\frac{1}{\sqrt{2^4}}\sum_{b_0=0}^1\sum_{b_1=0}^1\sum_{b_2=0}^1\sum_{b_3=0}^1
e^{-\frac{2\pi i(b_0+2b_1+4b_2+8b_3)}{16}}|b_0b_1b_2b_3\rangle
.
\]
Swapping qubits $3\leftrightarrow 0$ and $1\leftrightarrow 2$, and setting $|b\rangle=|b_3b_2b_1b_0\rangle$, gives
\[\frac{1}{\sqrt{2^4}}\sum_{b=0}^{2^4-1}e^{-\frac{2\pi i kb}{2^4}}|b\rangle=\QFT_{2^4}\ket{k},
\]
the quantum Fourier transform on $\ZZ_{2^4}$ applied to $\ket{k}\in \ZZ_{2^4}$.
This generalises in a straightforward manner to give the quantum Fourier transform of $\ZZ_{2^n}$, for arbitrary
$n$. Note that all the gates used are $2$--qubit gates.
In the general case the number of gates used is
$(n+1)n/2\le n^2$, as claimed.
\subsection{The continued fractions algorithm}\label{subs:cont_frac}
A {\em (finite) continued fraction} is an expression of the form
\[
a_0+\frac{1}{a_1+\frac{1}{a_2+\frac{1}{\cdots+\frac{1}{a_n}}}}
\]
where $a_0\in\ZZ$, $a_i\in\NN$, for each $i>0$, and $n\ge 0$. This finite continued fraction is
denoted $[a_0,\ldots,a_n]$ and clearly represents a unique rational number. Conversely,
using the Euclidean algorithm, it can be seen that a positive rational number can be
expressed uniquely as a finite continued fraction $[a_0,\ldots,a_n]$, with $a_n>1$
(see \cite{hardywright} or \cite{koblitz}).
\begin{ex}
As
\begin{align*}
125 & = 3\cdot 37 +14\\
37  & = 2\cdot 14 +9\\
14  & = 1\cdot 9 + 5\\
9 & = 1\cdot 5 +4\\
5 & = 1\cdot 4 +1
\end{align*}
we have
\begin{align*}
\frac{125}{37} & = 3\cdot\frac{37}{37}+\frac{14}{37} \\
& = 3+\frac{1}{\frac{37}{14}} \\
& = 3+\frac{1}{2\cdot\frac{14}{14}+\frac{9}{14}} \\
& = 3+\frac{1}{2+\frac{1}{\frac{14}{9}}} \\
&\qquad\vdots \\
& = 3+\frac{1}{2+\frac{1}{1+\frac{1}{1+\frac{1}{1+\frac{1}{4}}}}}
\end{align*}
and so the continued fraction representing ${125}/{37}$ is $[3,2,1,1,1,4]$.
\end{ex}

The {\em $j^{\rm th}$ convergent} of the continued fraction $[a_0,\ldots,a_n]$
is the expression $[a_0,\ldots,a_j]=p_j/q_j,$ say. Continuing the above example
the convergents of $125/37$ are $[3]=3$, $[3,2]=7/2$, $[3,2,1]=10/3$, $[3,2,1,1]=17/5$, $[3,2,1,1,1]=27/8$,
and $125/37$.
Clearly we may compute the convergents
of a given rational number using the Euclidean algorithm as in the above example and this allows
us to make use of Theorem \ref{th:contfrac} in Shor's algorithm.
The complexity of this algorithm, known as the {\em continued fraction algorithm}
is the same as that of the Euclidean algorithm: that is $O(L^3)$ operations are
required to compute the continued fraction of $p/q$, where $L=\max\{\log(p),\log(q)\}$.
\subsection{The hidden subgroup problem}\label{subs:hsp}
The problems of
factoring integers and finding the period of functions may
both be regarded as instances of the ``hidden subgroup problem'' which we discuss in this section.

Let $G$ be a group and $X$ a set and let $f:G\ra X$ be a function.
Assume that there is a subgroup $K\le G$ such that
\be[(i)]
\item
$f$ restricted to $gK$ is constant,
for all $g \in G$, and
\item if $gK\neq hK$ then $f(g)\neq f(h)$.
\ee
Then we say that $f$ is a {\em hidden subgroup function} which
{\em hides} the subgroup $K$ and that
$K$ is the {\em hidden subgroup} of $f$. The {\em hidden subgroup problem}
is to find the hidden subgroup of a given hidden subgroup function. That is, to solve the problem we must find
a generating set for $K$. (In some weaker versions of the problem it is only required that random elements
of $K$ are found.) We are concerned
here with the complexity of this problem and, in particular, whether or not it can
be solved more quickly using quantum rather than classical techniques.
We shall say that the hidden subgroup problem can be solved {\em efficiently}
if there is an algorithm which outputs generators of $K$ in time bounded by
some polynomial in $\log|G|$. This is a simplification, as in practice it is
necessary to consider how $G$ is encoded and the effect of this encoding on
the complexity of the problem. However for the purposes of the present
brief discussion it is enough to assume that our
quantum system has access to elements of $G$ in some appropriate form.

First of all we point out that the hidden subgroup problem has a number
of intrinsically interesting special cases.
For example, a periodic function $f$ with period $r$, in the sense of Section \ref{subs:overview},
on a cyclic group $C=\sgp{x}$, hides the subgroup $\sgp{x^r}$.
Therefore, period finding,
and hence factoring of integers
are, as claimed, particular cases of this problem.

Another problem which
may be described in this way is that of finding discrete logarithms.
Given a cyclic group $G$ of order $n$ generated by an element $g$ the
{\em discrete logarithm problem} is, given $a\in G$, to find the least
positive integer $r$ such that $a=g^r$. To formulate this as a hidden subgroup
problem consider the function $f:\ZZ_n\times \ZZ_n \ra \ZZ_n$ given by
$f(x,y)=g^x a^{-y}$. This is a homomorphism with kernel the subgroup $K$ generated
by $(r,1)\in \ZZ_n\times \ZZ_n$. Therefore $f$ hides the subgroup $K$. Clearly finding
the generator of $K$ gives us $r$. In fact if we can obtain any element $(s,t)\in K$ then
we can compute $r=s/t$ (in time $O(\log^2(n))$)
as long as we know that $K$ is generated by an element  of the form $(r,1)$.
The U.S. Digital Signature Algorithm is based on the assumption that no
polynomial algorithm is known for the discrete logarithm problem \cite{menezesoorschot}.
Details of an efficient quantum algorithm for this problem may be found in \cite{nielsenchuang}.

As a third example consider a group $G$ acting on a set $X$. If
$x\in X$ then we may define a function $f:G\ra X$ by $f(g)=g\cdot x$,
for $g\in G$. Then $f$ hides the stabiliser of $x$.

As a further example
we mention that the graph isomorphism problem may be viewed
as a special case of the hidden subgroup problem (see \cite[Section 6]{joszahs}). No polynomial time
algorithm for the graph isomorphism
problem is known but on the other hand it is not known to be NP--complete.
(A problem is NP if there is a classical non--deterministic polynomial time algorithm for its solution.
A problem is NP--complete if every problem which is NP may be reduced, efficiently, to this problem.
See \cite{papadimitrou} for further details.)
As a problem
which seems likely to lie outside the class of problems classically
solvable in polynomial time and which also seems unlikely to
be NP--complete, the
graph isomorphism problem is a good test case for the power of quantum computation.
In its formulation as a hidden subgroup problem  it becomes a question of finding
a hidden subgroup of a permutation group (of degree twice the number of vertices
of the graphs in question). An efficient quantum algorithm for the
hidden subgroup problem in permutation groups would therefore give rise to an efficient quantum algorithm
for the graph isomorphism problem. However, at the time of writing, there are very
few non--Abelian groups for which polynomial time quantum algorithms for
the  hidden subgroup
problem have been found.

The first quantum algorithm for a hidden subgroup problem appears to
be Deutsch's algorithm, where the hidden subgroup of $\ZZ_2$ is either trivial
or the entire group. However the subject begins in earnest with Simon's algorithm,
\cite{simon94} and \cite{simon97}, for a restricted case of
the hidden subgroup problem in $\ZZ_2^n$.
Simon's algorithm uses the
Walsh--Hadamard
transform on $\ZZ_2^n$ to extract information from
a quantum state,
 but the hidden subgroup here must have order $2$.
Shor \cite{shor1} realised that it was possible to implement the quantum Fourier transform for $\ZZ_{2^n}$
and used it instead of the Walsh--Hadamard transform to generalise Simon's algorithm.
This resulted in the factorisation algorithm
described above.
Subsequently methods for implementing and applying the quantum Fourier transform to
a wider class of Abelian groups
were developed by a number of people including
Shor \cite{shor2}, Cleve \cite{cleve94}, Coppersmith \cite{coppersmith}
and Deutsch.
Kitaev implemented the quantum Fourier transform  \cite{kitaev95} for arbitrary finitely
generated Abelian groups
and used it to construct his {\em phase estimation} algorithm which
finds eigenvalues of unitary transformations. More precisely, Kitaev's phase estimation
algorithm, given a unitary transformation and one of its eigenvectors $\ket{u}$, will return
a value $\phi$, where $e^{2\pi i\phi}$ is the eigenvalue corresponding to $\ket{u}$ (see for
example \cite{nielsenchuang}). Kitaev used the phase estimation algorithm to solve the problem
of finding stabilisers, as described above, where $G$ is a finitely generated Abelian group, and showed how
this gives rise to efficient algorithms for factoring integers and for the discrete logarithm
problem. Mosca and Ekert \cite{moscaekert} have shown that the phase estimation algorithm can be used
to solve the general hidden subgroup
problem in finitely generated Abelian subgroups.

Moving away from Abelian groups
Ettinger and Hoyer construct an algorithm which solves the hidden subgroup problem in the finite dihedral
group $G$, using at most a polynomial (in $\log|G|$) number of
calls to the unitary transformation $U_f$ simulating $f$
(as in Section \ref{subs:create_periodic}). However their algorithm requires
exponential time to interpret the output. That is, the part of the
algorithm analagous to the
continued fractions post processing in  Shor's algorithm requires exponentially many operations.
Ettinger, Hoyer and Knill \cite{ettingerhoyer} generalise this to show that there is a quantum
algorithm for the hidden subgroup problem in an arbitrary finite group $G$ which requires
$O(\log|G|)$ calls
to 
$U_f$. 
However they do not give explicit implementation of the measurements required, and the post processing
part of the algorithm is again exponential.
P\"uschel, R\"otteler and Beth \cite{puschelrotteler}, \cite{rottelerbeth} have implemented the quantum Fourier transform for
the wreath product $\ZZ_2^n\wr \ZZ_2$ and hence solve the hidden subgroup problem efficiently in these groups. Hallgren, Russell and Ta-Shma
\cite{hallgrenrussell} have shown that the special case of the hidden
subgroup problem where $K$ is a normal subgroup of a finite group $G$ can
be solved efficiently on a quantum computer. Their algorithm 
uses the Fourier transform for
an arbitrary finite group to distinguish $K$.
Ivanyos, Magniez and Santha \cite{ivanyosmagniez} generalise \cite{rottelerbeth} by constructing
polynomial time quantum algorithms for the hidden subgroup problem in
specific finite groups: namely groups having small commutator subgroups
and groups which have an elementary Abelian normal $2$--subgroup of small index or with
cyclic factor group. (Here {\em small} means of order polynomial in $\log|G|$.)

Friedl, Ivanyos, Magniez, Santha and Sen \cite{friedlivanyos} complete and generalise much of the above by extending these
last results to
solvable groups satisfying the following condition on their commutator subgroups.
A finite Abelian  group $A$ is said to be
{\em smooth} if it is the direct sum  of an elementary Abelian $p$--group, for some prime $p$,  with
a group of order polynomial in $|A|$. A finite solvable group $G$ is said to be
{\em smoothly} solvable if the Abelian
factors of its derived series are smooth Abelian groups.
In \cite{friedlivanyos} efficient quantum algorithms are constructed for the hidden subgroup problem in
finite solvable groups which have smoothly solvable commutator
subgroups.
These include semidirect products of the form $\ZZ_p^k\rtimes \ZZ_2$, where $p$ is a prime
(which reduce to  finite dihedral groups of order $2p$ when $k=1$) and
groups of upper triangular
matrices of bounded dimension over a finite field. In fact
their work builds on quantum algorithms for solvable groups developed by Watrous \cite{watrous2} and
Cheung and Mosca \cite{cheungmosca}.

Moore, Rockmore, Russell and Schulman \cite{moorerockmore} show that $q$--hedral groups $\ZZ_p\rtimes \ZZ_q$ have
quantum efficiently solvable hidden subgroup problem, when $q=(p-1)/g(\log(p))$, for some
polynomial $g$. They also prove that quantum efficiency of the hidden subgroup problem is closed
under taking certain extensions, as follows. Suppose that there is an efficient quantum algorithm
for the hidden subgroup problem in the group $H$. Let $G$ be a group with normal subgroup $N$ such
that $G/N=H$ and $|N|=g(\log|H|)$, for some polynomial $g$. Then there is an efficient quantum
algorithm for the hidden subgroup problem in $G$.

Hallgren \cite{hallgren}  considers the problem of finding the period of a periodic function $f$ from
the group $\RR$ to a set $X$, where the period may be irrational. Hallgren's quantum algorithm runs in time
polynomial in an appropriatly defined input size. This gives rise to
efficient algorithms for a number of computational problems of algebraic number theory: the solution of Pell's equation, the pricipal ideal problem and determination
of the class group. No efficient classical algorithms are known for any of these problems.
(A full and self--contained exposition of Hallgren's
algorithm and it's application to number theoretic problems may be found in \cite{joszape}.)
In \cite{lomonacokauffman} Lomonaco and Kauffman consider the hidden subgroup problem in $\RR$ and various other
Abelian groups which are not finitely generated.
\newpage
\section{Grover's algorithm}
\subsection{Overview}\label{subs:over}
Grover's algorithm \cite{groverhaystack} operates in a quite different way to Shor's. The basic problem is an unstructured search: We are given an
$N$ element set $X$ and a map $P:X\ra\{0,1\}$ and are required to find $x\in X$ such that $P(x)=1$. We call
a value of $x$ such that $P(x)=1$ a {\em solution} to the search problem.
In the first four parts of this Section
we consider the algorithm for the case where we know that there are exactly $M\ge 1$ solutions.
Then, in Section \ref{subs:BHT}, we show how techniques developed by Brassard,
H\o yer and Tapp can be used to remove this constraint.
In Section \ref{subs:grovercircuit} we describe the algorithm and then
in Sections \ref{subs:IATM} and \ref{subs:groverrot}
we explain why it works.

No extra information is known about $P$,
we merely have an oracle to evaluate $P(x)$ for a given $x\in X$.
Classically, the best algorithm (exhaustive testing)
requires $N-M+1$ evaluations to find a solution $x$ with certainty, since the first $N-M$ elements
tested may be non--solutions.
Probabilistically, we would expect to find a result after $N/2M$ evaluations.
In contrast, Grover's quantum algorithm performs the search in time $O(\sqrt{N/M})$ on a quantum computer.

The idea of the algorithm is, roughly speaking, the following.
Suppose that $N$ has size $2^n$ and,
as before, we
prepare the standard superposition of all possible outputs (entangled with inputs):
\begin{equation}\nonumber
 \frac{1}{\sqrt{2^n}}\sum_{x=0}^{2^n-1}\ket{x}\tens\ket{P(x)}.
\end{equation}
We wish to find a state $\ket{x}\tens\ket{1}$ for some $x$. By direct measurement at this stage,
there is only a probability of
$M/\sqrt{2^n}$ of finding such a state. In the worst case when there is only one solution this
falls to $1/\sqrt{2^n}$.
The strategy is to increase the amplitude of
vectors of the form $\ket{x}\tens\ket{1}$ and decrease the amplitude of those of the form $\ket{x}\tens\ket{0}$,
until the state approximates
\begin{equation}\label{e:groverwish}
\frac{1}{\sqrt{M}}\sum_{i=1}^{M}\ket{x_i}\tens\ket{1},
\end{equation}
where the solution set is $\{x_1,\ldots,x_M\}$.
Measuring this altered state then gives a solution with high probability.

\subsection{A circuit for Grover's algorithm}\label{subs:grovercircuit}
Let us consider how this strategy may be carried out in practice.
We assume, for simplicity, that $N=2^n$, for some positive integer $n$ and that we know in advance
that there  are exactly $M$ solutions, where $M\ge 1$.
The algorithm uses the standard oracle $U_P$ for the function $P$ which, as in Section \ref{subs:parallel}
maps $\ket{x}\tens\ket{y}$ to $\ket{x}\tens\ket{P(x)\oplus y}$. Thus the
quantum system underlying the
algorithm
consists of a first register of $n$ qubits and second register, called the {\em oracle workspace},
of a single qubit.
As in the description of Deutsch's algorithm in Section \ref{subs:gDJ},
we begin with the state $\quzero^{\tens n}\tens\quone$ to which is
applied $W^n\tens W$ followed by $U_P$.
As in Section \ref{subs:gDJ}, the first register then contains
\begin{equation}\label{e:D(P)}
\frac{1}{\sqrt{2^n}}\sum_{x=0}^{2^n-1}(-1)^{P(x)}|x\rangle=\cD(P),
\end{equation}
as defined in
\eqref{e:deutschstate}.
Note that the oracle maps the state $\ket{x}\tens \ket{w}$
to $(-1)^{P(x)}\ket{x}\tens \ket{w}$, so we may regard the second register as unchanged and the
amplitude of the first register as multiplied by $-1$ if and only if $x$ is a solution.

\begin{figure}
\begin{center}
\psfrag{o}{oracle action}
\psfrag{i}{inversion about the mean}
\psfrag{a}{after $O(\sqrt{N/M})$ operations}
\includegraphics[width=3.5in]{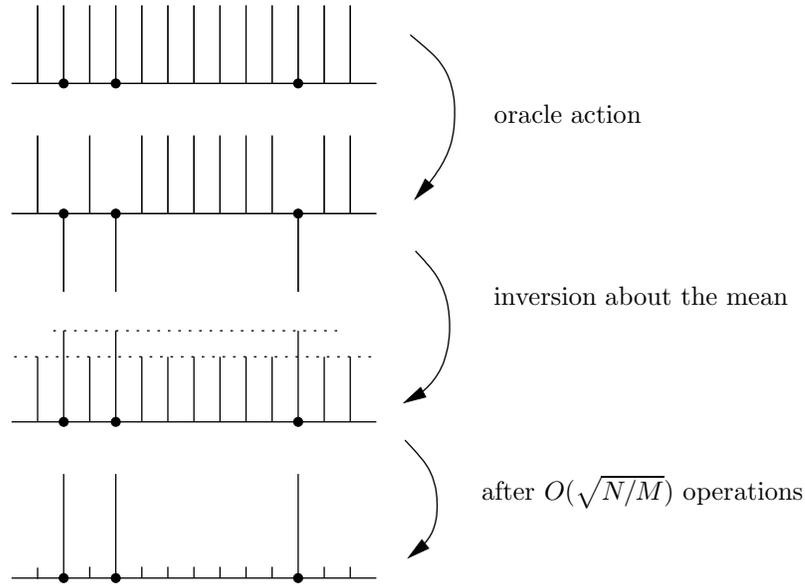}
\caption{\label{fig:grover1} Operation of Grover's algorithm}
\end{center}
\end{figure}

We now need to magnify the amplitudes of the vectors $\ket{x}$ where $x$ is a solution.
This is accomplished using
{\em inversion about the mean} which may be defined
as the
unitary transformation $F=W_nTW_n$, where $T$ is the {\em conditional phase shift} operator given by
\begin{equation}\nonumber
T\quzero = \quzero\quad\textrm{ and }\quad T\ket{x}=-\ket{x},\quad\textrm{ for all }\quad x\neq 0.
\end{equation}
We shall discuss inversion about the mean in more detail in Section \ref{subs:IATM}, for the time being
assuming that it does what we require of it: that is to increase negative amplitudes and
decrease positive ones.
With this assumption apply $F$ to the state $\cD(P)$ of \eqref{e:D(P)} (that is we apply $F\tens I$ to our quantum system). As shown in Figure \ref{fig:grover1} we then repeat the process, applying $U_P$ followed
by $F$ untill the amplitudes of the solutions approach $1/\sqrt{M}$ and the amplitudes of all other
basis vectors approach zero, as in \eqref{e:groverwish}.
We call the composite function
$\cG=(F\tens I)\circ U_P$ the  {\em Grover operator}.
The question is how do we know how many iterations of $\cG$
to allow before halting. It will become apparent in
Section \ref{subs:groverrot}, where we answer this question,
that we must choose the number $R$ of iterations carefully, as the amplitudes of the solutions
do not approach a steady state but rather oscillate, so too many iterations will be as bad as too few.
As we shall see, the required number of iterations is $O(\sqrt{N/M})$.

\begin{figure}[!btph]
\begin{center}
\psfrag{W}{$W^{\tens n}$}
\psfrag{nq}{$n$ qubits}
\psfrag{oracle}{oracle}
\psfrag{workspace}{workspace}
\psfrag{G}{$\cG$}
\psfrag{o}{$O(\sqrt{N/M})$ copies}
\includegraphics[width=4in]{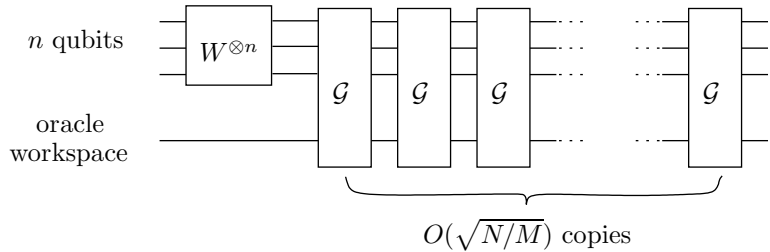}
\caption{\label{fig:grover2} The quantum circuit for Grover's algorithm}
\end{center}
\end{figure}

\begin{figure}[!btph]
\begin{center}
\psfrag{G}{$G$}
\psfrag{oracle}{oracle}
\psfrag{W}{$W^{\tens n}$}
\psfrag{x}{$\ket{x}\mapsto$}
\psfrag{p}{$(-1)^{P(x)}\ket{x}$}
\psfrag{phase}{phase shift $T$}
\psfrag{0}{$\ket{0}\mapsto\ket{0}$}
\psfrag{xm}{$\ket{x}\mapsto -\ket{x}$ if $x\not=0$}
\includegraphics[width=5in]{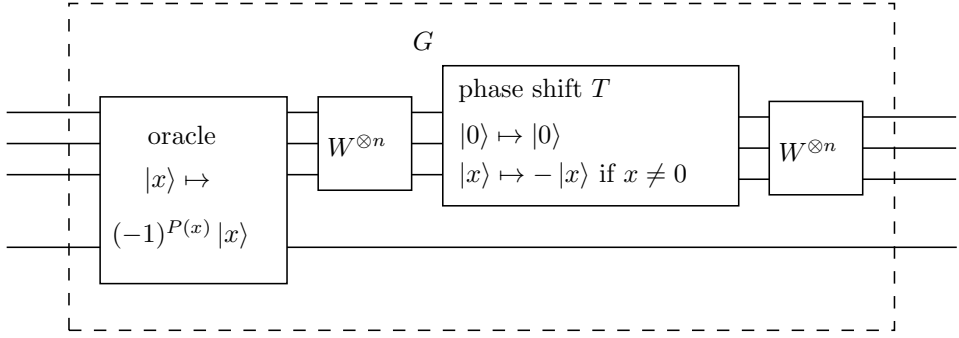}
\caption{\label{fig:grover3} Decomposition of the Grover operator}
\end{center}
\end{figure}

Grover's algorithm can be depicted using the circuit diagrams in Figures~\ref{fig:grover2} and \ref{fig:grover3}.

\subsection{Inversion about the mean}\label{subs:IATM}
We now explain why the operation $F$ of Section \ref{subs:over} is called ``inversion about the mean''
and behaves as shown in Figure \ref{fig:grover1}.
It is easy to verify that the conditional phase shift operator $T$ satisfies
\[T= 2|0\rangle\langle0|-I.\]
Inversion about the mean is then given by
\begin{equation}\label{e:IATM2}
 F=W^{\tens n}(2|0\rangle\langle 0|-I)W^{\tens n}.
\end{equation}
If we let
\[ |\psi\rangle=W^{\tens n}|0\rangle=\frac{1}{\sqrt{N}}\sum_{x=0}^{N-1}|x\rangle \]
then we have
\[ \langle \psi |=\langle 0| W^{\tens n}=\frac{1}{\sqrt{N}}\sum_{x=0}^{N-1}\langle x|. \]
It follows, from \eqref{e:IATM2}, that
\begin{equation}\label{e:IATM3}
F = 2|\psi\rangle \langle\psi|-I.
\end{equation}
Now consider the action of this operator on a general quantum state.
We have
\[
\left(2|\psi\rangle \langle\psi|-I\right)\sum_{k}\alpha_k|k\rangle  =
2\sum_k\alpha_k|\psi\rangle\langle\psi|k\rangle-\sum_k\alpha_k|k\rangle.
\]
Now,
\begin{eqnarray*}
\sum_k\alpha_k|\psi\rangle\langle\psi|k\rangle & = &
\sum_{k=0}^{N-1}\alpha_k\cdot\frac{1}{\sqrt{N}}\sum_{x=0}^{N-1}\langle x|k\rangle |\psi\rangle \\
& = & \frac{1}{\sqrt{N}}\sum_{k=0}^{N-1}\alpha_k|\psi\rangle \\
& = & \frac{1}{\sqrt{N}}\sum_k\alpha_k\frac{1}{\sqrt{N}}\sum_{x=0}^{N-1}|x\rangle \\
& = & A\sum_{x=0}^{N-1}|x\rangle,
\end{eqnarray*}
where
\[ A=\frac{1}{N}\sum_{k}\alpha_k \]
is the average (mean) of $\{\alpha_k\}$.
So
\[F \left( \sum_{k=0}^{N-1}\a_k\ket{k}\right)=\sum_{k=0}^{N-1}(2A-\alpha_k)|k\rangle. \]
Therefore $F$ acts by reflecting the amplitudes $\a_k$ about their mean value $A$.

\subsection{The Grover operator as a rotation}\label{subs:groverrot}
Suppose that there are $M\ge 1$ solutions in a search set of size $N$. Let
\[\begin{array}{ll}
\displaystyle  | a\rangle=\frac{1}{\sqrt{N-M}}\sum_{P(x)=0}|x\rangle, &
\displaystyle   | b\rangle=\frac{1}{\sqrt{M}}\sum_{P(x)=1}|x\rangle.
\end{array}
\]
The initial state of the system is
\[ |\psi\rangle=\frac{1}{\sqrt{N}}\sum_{x=0}^{N-1}|x\rangle,
\]
which can be written as
\begin{equation}\label{e:phiex}
\ket{\psi}=\sqrt{\frac{N-M}{N}}| a\rangle+\sqrt{\frac{M}{N}}| b\rangle.
\end{equation}
We claim that the Grover operator keeps the quantum state in the plane spanned by
$| a\rangle$ and $| b\rangle$, i.e.
that the subspace $S=\mbox{{\rm span}}\{| a\rangle,| b\rangle\}$ of the quantum system is invariant
under operation of $\cG$.
Since the oracle $U_P$ acts on both the first and second registers
it is convenient to define an operator
$O$ of the first register by $U_P(\ket{x}\tens\ket{w})=(O\ket{x})\tens\ket{w}$. This is possible since
$U_P$ leaves $\ket{w}$ unchanged. Thus $O$ determins the action of the oracle on the first register, given
that the second register is in state $\ket{w}$. Now let $G=F\circ O$. Then the operation of $\cG$ on
the first register is determined by $G$. We aim to show that $S$ is invariant under $G$.

We first consider the action of the oracle on $S$.
We have $O| a\rangle=| a\rangle$ and $O| b\rangle=-| b\rangle$.
Thus
\[O(\a | a\rangle+\b | b\rangle)=\a| a\rangle-\b\ket{ b}\in\mbox{{\rm span}}\set{\ket{ a},\ket{ b}},\]
so $S$ is invariant under the action of the oracle.
Geometrically, $O|_S$ is a reflection in the line through the origin defined by
$| a\rangle$ (by which we mean the set of points
$\a \ket{a}$, for $\a\in \CC$).

Next we consider the action of inversion about the mean, that is the operator
$F$, on $S$. From \eqref{e:IATM3} we have
\begin{eqnarray*}
F\left(\a | a\rangle+\b| b\rangle\right)
& = & 2\a\langle\psi| a\rangle|\psi\rangle+2\b\langle\psi| b\rangle|\psi\rangle-\a| a\rangle-\b| b\rangle.
\end{eqnarray*}
Since $|\psi\rangle\in S$ it follows that
$S$ is invariant under $F$ and, with the above, this implies that $S$ is invariant under $G$, as required.

\begin{figure}[!btph]
\begin{center}
\psfrag{a}{$\ket{ b}$}
\psfrag{b}{$\ket{ a}$}
\psfrag{p}{$\ket{\psi}$}
\psfrag{o}{$O\ket{\psi}$}
\psfrag{g}{$G\ket{\psi}$}
\psfrag{t}{$\theta$}
\psfrag{t2}{$\theta/2$}
\includegraphics[width=2in]{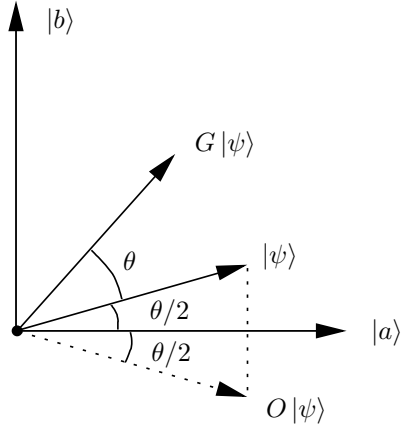}
\caption{\label{fig:grover4} Geometric interpretation of the Grover operator}
\end{center}
\end{figure}

Moreover, it is easy to see that $F|\psi\rangle=|\psi\rangle$ and that if
$\ket{\phi}$ is orthogonal to $|\psi\rangle$ then $F|\phi\rangle=-|\phi\rangle$.
Thus $F|_S$ is a reflection in the
line through the origin defined by $|\psi\rangle$.
Thus $G|_S$, being the composition of two reflections in lines through the origin, is a rotation
about the origin of the plane $S$. To find the angle of rotation note that since $1\le M\le N$ we
have $0\le\sqrt{N-M/N}< 1$, so that there exists $\theta\in \RR$ such that $0<\theta\le \pi$ and
$\cos(\theta/2)=\sqrt{N-M/N}$. From \eqref{e:phiex} we have therefore
\[\ket{\psi}= \cos \left(\frac{\theta}{2}\right)| a\rangle+\sin\left(\frac{\theta}{2}\right)| b\rangle. \]
Thus $G|_S$ is an
anticlockwise rotation
about the origin, through an angle $\theta$, as shown in Figure~\ref{fig:grover4}.
Hence
\[ G|\psi\rangle=\cos\left( \frac{3\theta}{2}\right)| a\rangle+\sin\left(\frac{3\theta}{2}\right)| b\rangle \]
and in general,
\[ G^k|\psi\rangle=\cos\left( \frac{2(k+1)\theta}{2}\right)| a\rangle+\sin\left(\frac{2(k+1)\theta}{2}\right)| b\rangle. \]
If we rotate $\ket{\psi}$ through $\cos^{-1}\left(\sqrt{M/N}\right)$ radians
then we obtain a state close to the desired vector $\ket{b}$.
Measuring this state we will, with high probability, observe $x$ such that $P(x)=1$, that is a solution
to the search problem.
Thus the
number of times we should iterate the Grover operator is given by
\[\displaystyle R=\left\lfloor\frac{\cos^{-1}\left(\sqrt{M/N}\right)}{\theta}\right\rfloor.
\]
If $M\lte N/2$ then $\theta/2\gte\sin\left(\theta/2\right)=\sqrt{M/N}$. Thus we obtain, in this case,
\[\frac{\pi}{4}\sqrt{\frac{N}{M}}\]
as  an upper
bound for $R$.
Note that if we iterate approximately $\pi\sqrt{N/M}/2$
times, then we have rotated back almost to $-| a\rangle$, and the probability of obtaining a
solution is much worse again. So determining the
appropriate number of iterations is a delicate matter.
In the case where $M=1$ as in Grover's original paper, the number of iterations
required is approximately $\pi\sqrt{N}/4$.

It can be shown that if $M\gte{N/2}$, i.e. more than half of the elements of the
search set are solutions, then the number of iterations required {\em increases} with $M$! (See \cite{nielsenchuang}.) However, if we know in advance that $M\ge N/2$ then sampling the set $X$ at random,
and checking for a solution using the oracle, we'll find a solution with probability at least $1/2$,
with only one call to the oracle.
Even when
it is not known in advance whether or not $M\ge N/2$,  by doubling the size of the first
register and padding it with non--solutions we can assume, at very low cost, that
in fact $M\le N/2$ and so use the above bound $R$ on the number of iterations required
(see \cite{nielsenchuang} for details). In conclusion the number of Grover operations
$\cG$ required for a solution to be found with high probability is $O(\sqrt{N/M})$.

We show in the next section, how we can estimate $M$ when it is not known in
advance.
Also, note the following.
\be[(i)]
\item
If the required probability of error is less than a given constant then
the oracle in Grover's algorithm can also be implemented in time $O(\sqrt{N})$ to search
an unstructured database. (See \cite{nielsenchuang}.)
\item The time complexity of
unstructured quantum search algorithms is known to be $\Omega(\sqrt{N})$
(see \cite{boyeretal}).
\ee
%
\subsection{The Brassard--H\o yer--Tapp counting algorithm}\label{subs:BHT}
In \cite{bht}, Brassard, H\o yer and Tapp describe general conditions under which Grover's techniques may be
used. They also give a novel method for approximating the number of values $x$ for which a boolean function $P$
satisfies $P(x)=1$. Some related work, on the eigenvalue analysis of the operator described in this section, is also due to Mosca.
A more developed version by all four authors above appears in \cite{bhmt}.
Given a system in state $\ket{\psi}$ the idea is to
find a superposition of $G^{m}\ket{\psi}$,
for all values of $m$ in a given range, where $\psi$ and $G$ are defined in Section \ref{subs:groverrot}.
Since $G$ is a rotation, the above state will be periodic in $m$.
So as in Shor's algorithm, we may apply the quantum Fourier transform of
$m$ which will find the period of the above state (efficiently).
>From this we can estimate the number of solutions (without actually finding any!)
and hence the number of rotations required for
Grover's algorithm above to find a solution with high probability.

\begin{figure}[!btph]
\begin{center}
\psfrag{me}{measure}
\psfrag{k0}{$\ket{0}$}
\psfrag{W}{$W$}
\psfrag{psi}{$\ket{\psi}$}
\psfrag{gpsi}{$G^{m}\ket{\psi}$}
\psfrag{C}{$C$}
\psfrag{m}{$\ket{m}$}
\psfrag{qftn}{$\QFT_R$}
\includegraphics[width=3in]{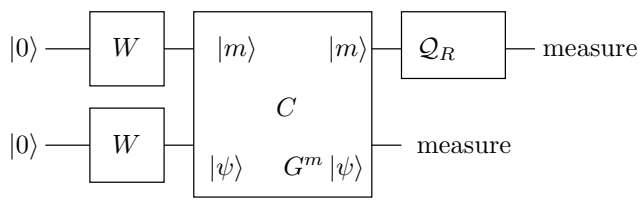}
\caption{\label{fig:bht} Brassard, H\o yer and Tapp's counting circuit}
\end{center}
\end{figure}

Given the Grover operator $G$ for $P$, define the {\em counting operator}
for $P$ to be
\[ C:\ket{m}\tens\ket{\psi}\mapsto\ket{m}\tens G^{m}\ket{\psi}. \]
Assume that the value $m$ can take any value in $\set{0,\ldots,R=2^r-1}$
Then the quantum circuit in Figure~\ref{fig:bht} will estimate the period of $C\circ (W\tens W)\ket{0}\tens\ket{0}$.
The accuracy to which we can estimate this period depends on how large a value of $R$ we take. More precisely, it is shown
in \cite{bht} (theorem 5) that if $t=|P^{-1}(1)|\lte N/2$ and $c$ is the output of this circuit, then
\[ |c-t|<\frac{2\pi}{R}\sqrt{tN}+\frac{\pi^2}{R^2}N \mbox{ with probability at least } \frac{8}{\pi^2}
.
\]

\newpage
\section{Watrous' algorithms for solvable groups}

Watrous \cite{watrous1},\cite{watrous2} has recently produced some group-theoretic work in quantum computing
which has quite a different nature to the hidden subgroup problem.
In particular, he describes a
Monte Carlo quantum algorithm which finds the order of a finite solvable group
in polynomial time. This uses some of the techniques of Shor's
algorithm, but also provides some new methods which depend critically on group
structure.

The group $G$ which is input to the algorithm is given as a finite black
box group \cite{babaiszemeredi}. This means that we assume the existence of some description of
$G$ as a set
of binary strings of fixed length $n$. Multiplication and inversion of
elements are performed, each at unit cost, by an oracle which knows
this description. The input to the algorithm is a finite generating
set for $G$, together with the oracle.

The representation of the elements of $G$ as binary strings gives us a
natural association between the elements of $G$ and a subset of the
basis elements of a $2^n$ dimensional vector space (that is, a register of $n$
qubits).
The notation $\ket{g}$ will be used to denote the basis element associated with the
group element $g$. One useful byproduct of Watrous' algorithm (and a vital
step within the algorithm) is the computation of a uniform superposition
\[ \frac{1}{\sqrt{|G|}}\sum_{g \in G} \ket{g} \] in one of the registers.

Classical Monte Carlo algorithms are already known which compute a
polycyclic generating set for a finite solvable group in polynomial
time \cite{babaicoopermanetal}. Hence we can assume that our starting
point for the algorithm is a generating set $g_1,\ldots g_k$ for which
the subgroups $H_j = \langle g_1,\ldots g_j\rangle$ form a subnormal series
(that is $H_j \lhd H_{j+1}$, for each $j$). In this case, each quotient group
$H_j/H_{j-1}$ is cyclic, of order  $r_j$ (we  define $H_0$ to be the identity
subgroup). The set of products of the form \[g_k^{a_k}\ldots g_1^{a_1},\]
with each $a_j$ ranging from 0 to $r_j-1$, provides a normal form, and
the group order is the product of the integers $r_j$. The problem is reduced
to finding the $r_j$, as the orders of the cyclic factor groups.

The algorithm works up the chain of the subgroups $H_j$, and so splits naturally
into $k$ steps. We shall describe just one such step, the $j$-th step, which
computes $r_j$.

\subsection{Step $j$ of the algorithm}
The $j$-th step of the algorithm splits into two phases. The first phase computes $r_j$, as the
period of the function
\[ f_j: \ZZ \rightarrow H_j/H_{j-1} \] defined by the rule
\[ f_j(a) = g_j^aH_{j-1} \]
using a fairly straightforward variant of Shor's algorithm. The second
phase, which involves some rather intricate calculation, uses knowledge
of the integer $r_j$ to compute a uniform superposition
of the elements of $H_j$. This superposition is then used as input
for the first phase of the next step. We shall describe both phases below.
To make the notation easier, from now on we shall abbreviate $r_j$ to $r$,
$g_j$ to $g$ and $H_{j-1}$ to $H$. In this case, $H_j$ is equal to $\sgp{g}H$.
Following Watrous we use the notation $\ket{H}$ for the superposition
\[\frac{1}{\sqrt{|H|}}\sum_{h \in H}\ket{h} \] and analogously
$\ket{\sgp{g}H}$ for the superposition over $\sgp{g}H$.

The computation takes place in a large tensor product space with
a number of different registers, of two
different types, which (like Watrous) we call $R$ registers and $A$ registers.
The $R$ registers are used to store superpositions of basis vectors indexed by
the elements of $G$. The $A$ registers are used to store superpositions of
basis vectors indexed by the integers in some finite range $0,\ldots M-1$
(We call the set of
such integers $\ZZ_M$. Note that this is a subset of $\ZZ$ which is not quite the
same as the cyclic group of integers mod $M$.) How big $M$ needs to be
differs between the two phases.

Recall that the quantum Fourier transform ${\cal Q}_M$ acts on an $A$ register as
\[ {\cal Q}_M: \ket{a} \mapsto \frac{1}{\sqrt{M}}\sum_{b \in \ZZ_M} e^{-2\pi iab/M} \ket{b}. \]
(In fact, Watrous uses the convention which calls this the inverse transform,
but for consistency we use the notation of the rest of this article.)
Group multiplication is provided by the unitary transformation
$U_G$ which acts on pairs of $R$ registers as
\[ U_G : \ket{g}\tens\ket{h} \mapsto \ket{g}\tens\ket{gh}. \]
A related unitary transformation $V_G^g$
acts on a pair of registers, one an $A$ register the other an $R$ register,
as
\[ V_G^g: \ket{a}\tens\ket{h} \mapsto \ket{a}\tens\ket{g^ah} \]
and is a vital ingredient to the variant of Shor's algorithm.

\subsection{The first phase of step $j$}
We shall describe this phase only briefly, aiming only to exhibit it as
a variant of Shor's algorithm, which is described in detail in Section
\ref{s:shor}.

Here the $A$ register needs to cover integers in the range $\ZZ_N$, where
$N$ is `large enough', basically $2^{2n+O(log1/\epsilon)}$, where $\epsilon$
is to bound the probability of error.
At the beginning of the $j$-th step the $A$ register contains $\ket{0}$,
and the $R$ register contains the uniform superposition $\ket{H}$.

We apply the inverse quantum Fourier transform ${\cal Q}_N^{-1}$ to the $A$ register,
then $V_G^g$ to the pair of registers, then the quantum Fourier transform
${\cal Q}_N$ to the $A$ register.

\begin{eqnarray*}
&&\ket{0}\tens\ket{H}\\
&&\\
&&\downarrow {\cal Q}_N^{-1}\tens I \\
&&\\
&&\frac{1}{\sqrt{N}} \sum _{a \in \ZZ_N}\ket{a}\tens\ket{H}\\
&&\\
&&\downarrow V_G^g \\
&&\\
&&\frac{1}{\sqrt{N}}\sum_{a \in \ZZ_N}\ket{a}\tens\ket{g^aH}\\
&&\\
&&\downarrow {\cal Q}_N\tens I  \\
&&\\
&&\frac{1}{N}\sum_{a \in \ZZ_N}\sum_{b \in \ZZ_N}e^{-2\pi iab/r}\ket{b}\tens\ket{g^aH}
\end{eqnarray*}
Observing $A$ gives some $b$ in $\ZZ_N$ which is (with high probability) a good
approximation for $\kappa/r_j$, for $\kappa$ random.

>From now on we follow Shor's standard procedure, as described in Section
\ref{s:shor}.
Using continued fractions, we can find integers $u,v$
such that $u/v=\kappa/r$, and with high probability $u$ and $v$ are coprime. We repeat this process an appropriate number of times to
give the required bound on the probability of error, and then find $r$ as
the lcm of the $v$ values.

\subsection{The second phase of step $j$}
The second phase has to extend the uniform superposition of
$H$ in the $R$ register
to a uniform superposition of $\sgp{g}H$, using the knowledge
of $r$ acquired
during the first phase.
The aim of this section is to explain that computation of
$\ket{\sgp{g}H}$ from $\ket{H}$.

In fact we produce several copies of $\ket{\sgp{g}H}$ from
several copies of $\ket{H}$. More precisely, we use $m=k-j+2$ $A$ registers
and $m$ $R$ registers to produce $m-1$ copies of
$\ket{\sgp{g}H}$ from $m$ copies of $\ket{H}$.
At the beginning of this calculation, each of the $m$ $R$ registers
contains the superposition $\ket{H}$. At the end,
$m-1$ of them contain $\ket{\sgp{g}H}$, and can be used in the next step, and the
other must be discarded.

In this phase the $A$ register is used to store the integers in $\ZZ_r$.
During the phase we use the inverse quantum Fourier transform ${\cal Q}_r^{-1}$,
the transformation $V_G^g$, and $U_G$.

The computation has two stages. During the first stage, we work with pairs
of registers, each pair consisting of
one $A$ register and one $R$ register. We do the same computation with each
pair, so to describe this stage we need only
say what happens to one such pair. At the beginning of the stage the $A$
register is set to $\ket{0}$
and the $R$ register to $\ket{H}$. First we apply the inverse quantum
Fourier transform ${\cal Q}_r^{-1}$ to the contents of the $A$ register
only, then we apply $V_G^g$ to the pair of registers, and then
${\cal Q}_r^{-1}$ again to the $A$ register. Then we observe the $A$
register, which projects onto the observed value. Denote the output state of this stage by $\ket{\psi}$. Then we have
\begin{eqnarray*}
&&\ket{0}\tens\ket{H}\\
&&\\
&&\downarrow {\cal Q}_r^{-1}\tens I  \\
&&\\
&&\frac{1}{\sqrt{r}} \sum _{a \in \ZZ_r}\ket{a}\tens\ket{H}\\
&&\\
&&\downarrow V_G^g \\
&&\\
&&\frac{1}{\sqrt{r}}\sum_{a \in \ZZ_r}\ket{a}\tens\ket{g^aH}\\
&&\\
&&\downarrow {\cal Q}_r^{-1} \tens I \\
&&\\
&&\frac{1}{r}\sum_{a \in \ZZ_r}\sum_{b \in \ZZ_r}e^{2\pi iab/r}\ket{b}\tens\ket{g^aH}\\
&&\\
&&\downarrow \textrm{measure }\,A\\
&&\\
&&\frac{1}{\sqrt{r}}\sum_{a \in \ZZ_r}e^{2\pi iab/r}\ket{g^aH} = \ket{\psi}
\end{eqnarray*}
The state $\ket{\psi}$ is almost what we want. It is precisely what we want if we struck lucky and observed $\ket{0}$ in the $A$
register. But otherwise it contains coefficients $e^{2\pi iab/r}$ which we would like to be able to replace by $1$'s.

What follows forms the crucial part of Watrous' argument. He shows that if we have superpositions $\ket{\psi}$ and $\ket{\psi'}$
in two distinct $R$ registers, then we can operate on the pair of registers in such a way that afterwards $\ket{\psi}$ is set to exactly
the superposition we want, and $\ket{\psi'}$ is unchanged (and hence it can be used again to `correct' a different $\ket{\psi}$, in
a different $R$ register). In fact we apply the operator $U_G$ to the pair of registers $c$ times in succession for a carefully
chosen value of $c$.
The notation $U_G^c$ is used to denote the composition of $U_G$, with itself,
$c$ times.

We suppose that \[ \ket{\psi} = \frac{1}{\sqrt{r}}\sum_{a \in \ZZ_r}e^{2\pi iab/r}\ket{g^aH},\quad
\ket{\psi'} = \frac{1}{\sqrt{r}}\sum_{a' \in \ZZ_r}e^{2\pi ia'b'/r}\ket{g^{a'}H}.\]
In order to calculate the effect of $U_G^c$ on $\ket{\psi}\tens\ket{\psi'}$, it helps first to consider
the effect on $\ket{\psi'}$ alone of the $R$ register operator $M_{g^ah}$, which is defined by the rule
\[ M_{g^ah}: \ket{x} \mapsto \ket{g^ahx}. \]
We see that
\begin{eqnarray*}
&&\ket{\psi'} = \frac{1}{\sqrt{r}}\sum_{a' \in \ZZ_r}e^{2\pi ia'b'/r}\ket{g^{a'}H}
= \frac{1}{\sqrt{r|H|}}\sum_{a' \in \ZZ_r} e^{2\pi ia'b'/r}\sum_{h' \in H}\ket{g^{a'}h'}\\
\\
&&\downarrow M_{g^ah} \\
\\
&& \frac{1}{\sqrt{r|H|}}\sum_{a' \in \ZZ_r} e^{2\pi ia'b'/r}\sum_{h' \in H} \ket{g^ahg^{a'}h'}\\
&=& \frac{1}{\sqrt{r|H|}}\sum _{a' \in \ZZ_r} e^{2\pi ia'b'/r}\sum _{h' \in H} \ket{g^{a+a'}h''h'}, \quad\mbox{for some $h''$
(since $H\lhd \sgp{g}H$)}\\
&=& \frac{1}{\sqrt{r|H|}} \sum _{a' \in \ZZ_r} e^{2\pi ia'b'/r}\sum _{h' \in H}
\ket{g^{a+a'}h'}\quad\mbox{(rewriting $h'$ for $h''h'$)} \\
&=& \frac{1}{\sqrt{r|H|}} e^{-2\pi iab'/r}\sum _{a' \in \ZZ_r} e^{2\pi i(a'+a)b'/r} \sum _{h' \in H}
\ket{g^{a'+a}h'} \\
&=& \frac{1}{\sqrt{r}} e^{-2\pi iab'/r}\sum _{a' \in \ZZ_r} e^{2\pi i(a'+a)b'/r} \ket{g^{a'+a}H} \\
&=& \frac{1}{\sqrt{r}} e^{-2\pi ab'/r}\sum _{a'' \in \ZZ_r} e^{2\pi ia''b'/r} \ket{g^{a''}H},
\quad\mbox{for $a''\in\ZZ_r$, $a''\equiv a'+a$ mod $r$}\\
&=&e^{-2\pi iab'/r}\ket{\psi'}.
\end{eqnarray*}
Note that in going from the third last to the second last line,
the equation $g^{a'+a}H = g^{a''}H$
follows from the fact $gH$ has order $r$ in
$\sgp{g}H/H$.
Now, since $U_G(\ket{g^ah}\tens \ket{h'})=\ket{g^ah}\tens M_{g^ah}\ket{h'}$,
\begin{eqnarray*}
  && \ket{\psi}\tens\ket{\psi'}=\frac{1}{\sqrt{r}}\sum_a e^{2\pi iab/r}\ket{g^aH}
  \tens\ket{\psi'}\\
  \\
  && \downarrow U_G\\
  \\
  && \frac{1}{\sqrt{r}} \sum_a e^{2\pi iab/r} \sum_h \ket{g^ah} \tens M_{g^ah}\ket{\psi'}\\
  &=&\frac{1}{\sqrt{r}} \sum_a e^{2\pi iab/r} \sum_h \ket{g^ah}\tens e^{-2\pi iab'/r}\ket{\psi'}\\
  &=& e^{-2\pi i a b' /r}\ket{\psi}\tens\ket{\psi'}.
\end{eqnarray*}
Hence, provided that $cb'=b$ mod $r$,

\begin{eqnarray*}
  && \ket{\psi}\tens\ket{\psi'}\\
  \\
  && \downarrow U_G^c\\
  \\
  &=& e^{-2\pi i a b' c/r}\ket{\psi}\tens\ket{\psi'}\\
  &=& e^{-2\pi i a b' /r}\ket{\psi}\tens\ket{\psi'}\\
  &=& \frac{1}{\sqrt{r}}\sum_{a \in \ZZ_r} \ket{g^aH}\tens \ket{\psi'}\\
  &=& \ket{\sgp{g}H}\tens \ket{\psi'}.
\end{eqnarray*}
The first of the two R-registers thus contains the state $\ket{\sgp{g}H}$.

\rule{0in}{.5in}
\begin{flushleft}
  {\small Michael Batty$^{\dagger}$,\\
    e-mail: Michael.Batty@ncl.ac.uk}
\end{flushleft}
\begin{flushleft}
  {\small Samuel L. Braunstein$^{\ddagger}$,\\
    e-mail: schmuel@cs.york.ac.uk}
\end{flushleft}
\begin{flushleft}
{\small Andrew J. Duncan$^{\dagger}$,\\
  e-mail: A.Duncan@ncl.ac.uk}
\end{flushleft}
\begin{flushleft}
  {\small Sarah Rees$^{\dagger}$,\\
    e-mail: Sarah.Rees@ncl.ac.uk}
\end{flushleft}
\begin{flushleft}
  {\small $\dagger$ Department of Mathematics,\\
    School Of Mathematics and Statistics,\\
    Merz Court,\\ University of Newcastle upon Tyne,\\
    Newcastle upon Tyne,\\ NE1 7RU,\\ UK.\\}
\end{flushleft}
\begin{flushleft}
  {\small $\ddagger$ Department of Computer Science,\\
    University of York,\\
    York,\\
    YO10 5DD,\\
    UK.}
\end{flushleft}

\end{document}